\documentclass[fleqn,10pt]{wlscirep}
\usepackage[utf8]{inputenc}
\usepackage[T1]{fontenc}
\usepackage{textgreek}
\usepackage{bm}
\title{Magnetic reconnection in the era of exascale computing and multiscale experiments}

\author[1,2,*]{Hantao Ji}
\author[3]{William Daughton}
\author[2]{Jonathan Jara-Almonte}
\author[3]{Ari Le}
\author[3]{Adam Stanier}
\author[2]{Jongsoo Yoo}
\affil[1]{Department of Astrophysical Sciences, Princeton University, Princeton, U.S.A.}
\affil[2]{Princeton Plasma Physics Laboratory, Princeton University, Princeton, U.S.A.}
\affil[3]{Los Alamos National Laboratory, Los Alamos, U.S.A.}

\affil[*]{e-mail: hji@pppl.gov}

\begin{abstract}
Astrophysical plasmas have the remarkable ability to preserve magnetic topology, which inevitably gives rise to the accumulation of magnetic energy within stressed regions including current sheets.  This stored energy is often released explosively through the process of magnetic reconnection, which produces a reconfiguration of the magnetic field, along with high-speed flows,  thermal heating, and nonthermal particle acceleration.  Either collisional or kinetic dissipation mechanisms are required to overcome the topological constraints, both of which have been predicted by theory and validated with in situ spacecraft observations or laboratory experiments. However, major challenges remain in understanding magnetic reconnection in large systems, such as the solar corona, where the collisionality is weak and the kinetic scales are vanishingly small in comparison to macroscopic scales. The plasmoid instability or formation of multiple plasmoids in long reconnecting current sheets is one possible multiscale solution for bridging this vast range of scales, and new laboratory experiments are poised to study these regimes.  In conjunction with these efforts, we anticipate that the coming era of exascale computing, together with the next generation of observational capabilities, will enable new progress on a range of challenging problems, including the energy build-up and onset of reconnection, partially ionized regimes, the influence of magnetic turbulence, and particle acceleration.
\end{abstract}

\begin{document}

\flushbottom
\maketitle
\thispagestyle{empty}

\noindent \textbf{Website summary:} Magnetic reconnection explosively releases stored magnetic energy in astrophysical plasmas. Thanks to advances in observations, exascale computing and multiscale experiments, it will be possible to solve outstanding physics problems, including the immense separation between global and dissipation scales, reconnection onset, and particle acceleration.

\vspace{5mm}

\noindent \textbf{Key points:} 
\begin{itemize}
\item Major challenges remain in understanding magnetic reconnection in large astrophysical systems where dissipation scales are extremely small compared to macroscopic scales.
\item The plasmoid instability of reconnecting current sheets is a natural mechanism to bridge this vast range of scales in both fully- and partially-ionized plasmas.  
\item Upcoming multiscale laboratory experiments are poised to provide the first validation tests of the plasmoid instability, whereas exascale simulations will allow researchers to evaluate competing hypotheses regarding the influence of turbulence.
\item These simulations and experiments can also shed new light on the mechanisms of reconnection onset, and how the reconnection layers couple with the macroscale systems that supply the magnetic flux.
\item Rapid progress is being made towards understanding the acceleration of highly energetic particles produced by magnetic reconnection, which may have broad relevance to energetic phenomena across the Universe.
\end{itemize}

\vspace{5mm}

\label{sec:intro}
\noindent
Most of the visible Universe exists in the plasma state, often accompanied by dynamically important magnetic fields produced within the plasma or arising from nearby compact objects.  Since plasmas are excellent electrical conductors, the magnetic field is nearly `frozen-in' into the plasma motion \cite[Chap.~2]{birnpriest:2007}. In the limit of infinite conductivity, this fundamental organizing concept is often referred to as `ideal evolution' and permits the labeling of field lines within a plasma, perfectly preserving the magnetic topology (connectivity of field lines).  Because most systems feature plasma flows and complex interactions between magnetic structures over a range of scales, the field lines invariably become tangled and develop highly stressed regions with strong `magnetic shear' ($ |\bm{\nabla B}| \gg B/L$, where $\bm{B}$ is magnetic field and $L$ is the global length scale). At some point in the evolution, these stressed fields rapidly change connectivity due to a process called magnetic reconnection \cite{priest00,yamada10,yamada22}.  Some significant fraction of the stored magnetic energy is converted to particle kinetic energy, including high-speed flows, thermal heating and, in many cases, nonthermal particle acceleration.   Magnetic reconnection is ubiquitous in space, laboratory and a growing number of astrophysical applications surveyed in ref. \cite{ji:2011}.  Familiar examples include the Earth's magnetosphere, the solar atmosphere and laboratory fusion plasmas.   Within astrophysics, reconnection may be part of the underlying processes generating some of the most energetic particles in the Universe, including the recently discovered Fast Radio Bursts~\cite{Zhang:2020}, and may even influence the habitability conditions for life on exoplanets \cite{MacGregor2021}.  Some classic examples are illustrated in Fig.~\ref{Example}, including the locations with high magnetic shear, which in the simplest form correspond to electric current sheets. 

Unlike many monographs on the subject, this Roadmap does not intend to be comprehensive, but forward-looking, aiming at outlining the new directions enabled by new research capabilities in exascale computing, multiscale experiments, and next-generation observations, to solve major open physics problems for magnetic reconnection in large systems. Below we begin with an overview of the problem and a brief summary of relevant history, followed by descriptions of frontier research opportunities respectively in multiscale physics, reconnection onset, and particle acceleration and heating.

\section*{Overview}
There are several key elements for understanding the physics of reconnection at a deeper level. First, the frozen-in properties in an electron-ion plasma are associated with the electron fluid due to its light weight. This is expressed in terms of the time changing rate of magnetic flux ($\psi \equiv \int \bm{B} \cdot d \bm{A}$) through an arbitrary area, $\bm{A}$ (enclosed by loop $\ell$), convecting with the electron flow as $d \psi/dt =  \oint  \left( \bm{E} + \bm{V}_e \times \bm{ B} \right) \cdot d \ell=0$ where $\bm{E}$ is electric field and $\bm{V}_e$ is the electron fluid velocity. Thus, the frozen-in condition is regulated by the electron momentum equation (the generalized Ohm's law)
\begin{equation}
\label{OhmsLaw}
\underbrace{ \bm{E} + \bm{V}_e \times \bm{B} }_{Ideal} \;\: = \; \underbrace{\bm{R}_{coll}}_{Collisional}  - \underbrace{\frac{\bm{\nabla} \cdot \bm{P}_e}{en} - \frac{m_e}{e} \frac{d \bm{V}_e}{dt}}_{Kinetic}  
\end{equation}
where $n$, $\bm{P}_e$, $m_e$ and $e$ are the electron number density, pressure tensor and mass, and elementary charge, respectively, and $\bm{R}_{coll}$ is the collisional force per electron per electron charge. In a fully ionized collisional plasma, $\bm{R}_{coll} \approx \eta \bm{j}$ where $\eta$ is the resistivity due to Coulomb collisions and $\bm{j}$ is the electric current density. If the terms on the right hand side of Eq.~\eqref{OhmsLaw} are negligible, then $d\psi/dt\approx 0$ and the magnetic flux is `frozen-in' to the electron flow. For the generic reconnection layer illustrated in the middle panel of Fig.~\ref{Overview}, deviations from ideal evolution occur within the `diffusion region' (blue), where either finite resistivity or kinetic effects (electron inertia and pressure tensor) are important.  Within the diffusion region,  field lines converging from opposite sides of the layer can change connectivity. 

The next key element for understanding the physics of magnetic reconnection is to grasp the remarkable global consequences of changing field line connectivity within a localized region. In particular, the newly reconnected field lines have a large curvature ($\bm{B} \cdot \bm{\nabla}\bm{B}$) which produces a tension force -- closely analogous to a stretched rubber band.  Allowing for the possibility of pressure anisotropy, the magnetohydrodynamic (MHD) momentum equation perpendicular to magnetic field is 
\begin{equation}
\label{ionMomentum}
\rho \left. \frac{d\bm{V}}{dt} \right|_\perp \approx \underbrace{- \nabla_\perp \left( P_\perp + \frac{B^2}{2 \mu_0} \right)}_{Pressure}  + \underbrace{ \left(1 + \frac{P_\perp - P_\|}{B^2/\mu_0}\right) \bm{B} \cdot \bm{\nabla} \bm{B}}_{Magnetic\; tension}\;\; - \underbrace{\left. \bm{F}_{coll}\right|_\perp}_{Collisional} 
\end{equation}
where $\rho$ is mass density, $\bm{V}$ is the fluid velocity, $\mu_0$ is vacuum permeability, $P_\perp$ ($P_\|$) is the total plasma pressure perpendicular (parallel) to the magnetic field, and $\bm{F}_{coll}$ is due to collisions between ions (viscous force) and between ions and neutral particles (frictional force). For newly reconnected field lines, the magnetic tension in Eq.~\eqref{ionMomentum} drives an outflow jet approaching the Alfv\'en speed $V_{out} \approx V_A \equiv B_0/\sqrt{\mu_0 \rho}$ ($B_0$ is the reconnecting field component).  The resulting deficit in magnetic pressure pushes new field lines into the diffusion region with a maximum inflow velocity $V_{in}\sim (0.01 \rightarrow 0.1)V_A$ (see below for details).  As this process continues, the larger stressed region is relaxed, leading to a reconfiguration of the global magnetic field on fast Alfv\'enic time scales $\sim L/V_A$. Assuming the diffusion regions remain small in comparison to the global scales, most plasma enters the flow jet across the magnetic separatrices as illustrated in Fig.~\ref{Overview}.  This inflow is a consequence of changing the field line connectivity within the diffusion region, which causes the entire extent of the reconnected field lines to join the outflow.  In this limit, the majority of the energy release is associated with the relaxation of field-line tension within outflow jets over long distances.  Since the spatial extent of these jets is limited only by the macroscopic configuration, they are one of the most prominent signatures of reconnection in both in-situ and remote-sensing observations.   For this Roadmap, our primary focus is on situations where the available magnetic energy to drive reconnection is comparable or larger than the initial plasma thermal energy, corresponding to $\beta  \equiv P /(B_0^2/2\mu_0) \lesssim 1$.  In these regimes, global relaxation of field-line tension is the ultimate `engine' for reconnection and is essentially `ideal', thus operating in a similar manner for most applications, but with a few important exceptions.   Plasma heating preferentially along the magnetic field ($P_\| \gg P_\perp$) can weaken the magnetic tension force in Eq.~\eqref{ionMomentum}, whereas in partially ionized regimes the jet formation is more complicated due to interactions with neutrals.  In addition, non-ideal kinetic physics may persist along magnetic separatrices \cite{lapenta15} (see red lines in Fig.~\ref{Overview}) to larger distances, while in very large systems shocks may form along these boundaries \cite{zhang19} and play a role in the energy conversion.   

In contrast to the ideal physics driving the jet, the non-ideal terms within the  diffusion region are intimately dependent upon the plasma conditions and spatial scales, and thus, a variety of different regimes are possible, as illustrated in the various panels of Fig.~\ref{Overview}.   Since the outflow is always energetically limited to $V_A$ in a quasi-steady state, mass conservation implies that the geometry of the diffusion region determines the dimensionless reconnection rate, $R \equiv V_{in}/V_{A} \approx \Delta/L$ where $\Delta$ is the diffusion region thickness. The current understanding of the diffusion region physics has evolved over more than 60 years in three main stages, yet a full understanding remains elusive for large space and astrophysical problems.  

\subsection*{\label{sec:history}History of magnetic reconnection in three acts}
The first stage of research began within the MHD description of the plasma, the simplest asymptotic theory which is valid when the diffusion region thickness ($\Delta$) remains larger than the relevant kinetic scale, either ion sound radius ($\rho_s$) or ion skin depth ($d_i$). Here $d_i\equiv c/\omega_{pi}$ where $c$ is speed of light and $\omega_{pi}$ is ion plasma frequency. $\rho_s \equiv \sqrt{(T_i+T_e) m_i}/q_i B$ where $T_e$ and $T_i$ are electron and ion temperatures, $B$ is magnetic field, and $m_i$ and $q_i$ are the ion mass and charge. The earliest model by Peter Sweet and Eugene Parker \cite{parker57a} (SP model) predicts that the reconnection rate scales as $R \approx 1/\sqrt{S}$ where $S\equiv \mu_0 V_A L/\eta$ is the Lundquist number.  Since for many applications~\cite{ji:2011} $S > 10^{8} \rightarrow 10^{30}$, the model implies a large aspect ratio $L/\Delta = \sqrt{S}$ diffusion region, which effectively chokes-off the inflow and thus limits the reconnection rate to values far smaller than observations. 

This failure motivated the Petschek model~\cite{petschek64a}, as well as more generalized steady-state models of fast reconnection in resistive-MHD~\cite{priestforbes1986,forbes87}, in which the length of the layer can be significantly shorter than the global scale. In the Petschek model, standing slow-mode shocks emanating out from a microscale length diffusion region allow the reconnection rate to reach a maximum value of order, $R\sim 0.01-0.1$. The basic ideas were brilliant, however, numerical simulations performed many years later found that such solutions required localized anomalous (possibly unphysical) resistivity profiles to be physically realizable, whereas layers tended to elongate and become SP-like for a uniform (or Spitzer) resistivity profile~\cite[and references therein]{forbes:2013} in agreement with both theory and laboratory experiments (the top left panel~\cite{ji98} in Fig.\ref{Overview}). Although Petschek-type solutions continue to attract research interest in solar physics, evidence for the required anomalous resistivity has not been found in laboratory experiments, magnetospheric observations~\cite{Torbert:2016} or kinetic simulations~\cite{roytershteyn12,liu:2013,le:2018} in plasmas with $T_i \gtrsim T_e$. However, this mechanism may still be important in some situations, see Box 1 for some kinetic instabilities which could lead to enhanced resistivity or viscosity. Nevertheless, the shocks envisioned by Petschek may occur along sufficiently large outflow jets. This might happen if the diffusion region is localized by kinetic effects, or possibly through more complex dynamic scenarios \cite{shibayama:2019}. In either case, these shocks may play a role in the global dynamics and heating\cite{zhang19}.

For a period of time, it was thought that fast reconnection may require physics beyond the MHD description, partly motivated by the observed fast reconnection in laboratory tokamak fusion plasmas~\cite{vongoeler:1974}.  The next major stage began with the realization that fast reconnection is easily achieved for kinetic-scale diffusion regions.    This can occur when the thickness of a resistive SP layer falls below the relevant ion scale ($\Delta_{SP} \equiv L /\sqrt{S} \lesssim d_i,\rho_s$) leading to rapid changes in the diffusion region dynamics \cite{aydemir92,wang93,kleva95}, and resulting in a reconnection rate $R \sim 0.1$ consistent with both two-fluid \cite{aydemir92,wang93,kleva95,birn01a,rogers01a} and kinetic simulations \cite{hesse01b,pritchett01} and also laboratory experiments \cite{ren05a,fox17} (the top middle panel~\cite{yoo13} in Fig.\ref{Overview}). A large body of work has confirmed these basic results, including many first principles fully kinetic simulations which are in beautiful agreement with in-situ spacecraft measurements in the Earth's magnetosphere \cite{Burch2016,Chen2017,egedal2019prl} as well as with well-diagnosed laboratory experiments~\cite{ji08,yamada18}. Although this experimental validation represents a significant milestone in our understanding of kinetic reconnection, there remain a range of unsolved problems.  Most notably, there is still no analytic theory that fully accounts for the nonlinear structure of a kinetic diffusion region and the associated reconnection rate.  A variety of ideas\cite{cassak2017} have been proposed, including dispersive waves\cite{rogers01a},
pressure tensor effects\cite{bessho05a,ng:2011}, plasmoids or magnetic islands\cite{daughton06a}, and coupling to the larger MHD scales\cite{wang01,Liu2017}, but a full understanding remains elusive.  Without a rigorous theory, it remains unknown how kinetic reconnection will scale to larger systems.   Whereas it is well-established \cite{Liu2017,cassak2017}, that local reconnection rates for kinetic-scale layers are typically $R\sim0.1$, it not clear how this connects to the global rate in dynamically evolving systems.  For example, some studies have demonstrated a strong system size dependence~\cite{stanier:2015, ng:2015} when the reconnection layer forms dynamically due to the merging of large magnetic islands.   It appears that the pressure anisotropy that develops during reconnection can lead to modifications to the field-line tension over macroscopic scales, as expressed in Eq.~\eqref{ionMomentum}. The ability to capture these effects within two-fluid models~\cite{allmann:2018,ng:2017} is crucial for developing global models of the Earth's magnetosphere and may likely be important even for the solar corona \cite{arnold:2021}. 

The third major stage of research began with the realization that the steady-state SP model is dynamically unstable in large astrophysical plasmas, offering a possible mechanism for the multiscale coupling~\cite{tajima:1997,shibata:2001}.  In particular, from linear tearing theory, SP layers are structurally unstable~\cite{birn80,biskamp82,forbes83,lee86,tajima:1997,shibata:2001,loureiro07,bhattacharjee09,samtaney2009,ni:2010,huang:2011,comisso:2016} to the formation of secondary magnetic islands (plasmoids) when $S>S_c$, where $S_c$ is the critical value determined by the stabilizing influence of flow shear~\cite{bulanov78} and the convection time through the layer.  Although not predicted by linear theory, numerical simulations \cite{bhattacharjee09,samtaney2009}, begin to feature plasmoids for $S_c \approx 10^4$, corresponding to diffusion regions with $L/\Delta \approx \sqrt{S_c} \sim 100$.   Whereas the existence of this instability was known in earlier work \cite{biskamp82,forbes83}, its significance was not appreciated due to the low Lundquist numbers used in the simulations.  However, the instability becomes increasingly violent \cite{tajima:1997,loureiro07} for larger Lundquist numbers, leading to a breakup of the layer into a large number of plasmoids, with new current sheets between each.  New current sheets can also be plasmoid unstable provided that $S>S_c$, where $S$ is based on the new, local parameters of each sheet.  In this plasmoid regime, MHD simulations~\cite{bhattacharjee09,loureiro12,huang:2011} and theoretical models~\cite{uzdensky10} predict that the reconnection rate is independent of $S$ with $R\sim 0.01 \sim 1/\sqrt{S_c}$ consistent with the critical aspect ratio for plasmoid formation.  Within this plasmoid mediated regime, the thickness of each new layer scales as $\Delta \propto \sqrt{\eta L}$ , where $L$ is the length of the parent current sheet, leading to a rapid downward hierarchy of scales \cite{shibata:2001} as illustrated in the lower left panel of Fig.~\ref{Overview}.   Presumably, this downward cascade will only be stopped if the new layer is stable, or if the thickness approaches the ion kinetic scale, at which point kinetic reconnection will be triggered.   This scenario has been confirmed using fully kinetic simulations with a Monte-Carlo treatment of the collision operator in both 2D (refs \cite{daughton09a,daughton09b}) and in 3D (ref.\cite{stanier:2019}) where the plasmoids form extended 3D flux ropes, which can interact in complex ways not possible in 2D (see Box 1).  Both kinetic and MHD simulations indicate that turbulence is naturally self-generated within 3D reconnection layers due to this flux rope dynamics.   

\subsection*{Frontier research opportunities}

The emerging picture of plasmoid-mediated reconnection offers an appealing mechanism for bridging MHD and kinetic scale dynamics.  If correct, these ideas would allow researchers to predict how reconnection may proceed in large astrophysical plasmas, which are well beyond the limitations of foreseeable computers to include the full range of scales.  In particular, our current understanding can be conveniently summarized in terms of reconnection phase diagrams in both fully and partially ionized regimes.  Although portions of these phase diagrams are experimentally validated, neither the plasmoid-mediated collisional regime nor the plasmoid-induced transition from collisional to kinetic scales has been tested experimentally. As described in the next section, this has motivated a new generation of laboratory experiments along with high-fidelity simulation tools to model these experiments.  

Looking ahead, perhaps the largest uncertainty in the phase diagram is the influence of MHD turbulence within asymptotically large reconnection layers.  As an alternative to the plasmoid picture, researchers have long argued that MHD turbulence may offer an explanation for fast reconnection in many astrophysical systems \cite{lazarian99}.   As illustrated in the lower right panel of Fig.~\ref{Overview}, this model argues for the existence of a `turbulent diffusion region' in which the thickness ($\Delta \gg d_i,\rho_s$) of the layer and the reconnection rate are both controlled by the properties of the turbulence.   Within this turbulent diffusion region, there would exist many smaller reconnection layers that could possibly extend down to kinetic scales (somewhat similar to the lower left panel, but without the dominance of plasmoids).   In contrast to plasmoid mediated reconnection, the turbulence is thought to control the thickness of the diffusion region, whereas new plasmoids regulate the length.  However, there are many assumptions in the proposed model \cite{lazarian99} regarding the MHD turbulence (externally forced or injected, isotropic, weak at injection length scale) which may not be generally valid.  As discussed later, there is presently no clear consensus regarding how reconnection may proceed in thicker turbulent layers.  With ongoing advances in computing, we believe there are opportunities for progress towards understanding the influence of turbulence in both MHD and kinetic regimes.  Testing these ideas in laboratory experiments will require new techniques for driving magnetic turbulence.

One crucially important aspect that is not represented in the phase diagram is the build-up and onset of explosive reconnection events.  For various applications, there often exist several promising and competing candidate scenarios to trigger reconnection, but generally they fall into one of two categories: external drive or macroscopic instabilities, as described later. The former category refers to the case when, even under constant external drive, the onset moment of fast reconnection is determined by internal multiscale dynamics within the reconnecting current sheets. In contrast, in the latter category the driving free energy is stored in places such as large-scale flux ropes which can be destabilized to trigger reconnection.  Trigger scenarios in each of these categories can be tested and challenged in proper numerical and laboratory settings.

Particle acceleration is one of newest areas where magnetic reconnection is frequently invoked to explain many of the energetic phenomena observed in our solar system and across the Universe.  Although a variety of ideas have been proposed, one of the most promising is a Fermi-type acceleration process arising from the curvature drift of particles within the relaxing magnetic field lines. Under certain conditions, this mechanism within simulations leads to power-law distributions, as often implied by observations. These mechanisms are summarized in a later section where their global implications are also discussed. Rapid progress based on numerical simulations is expected in this area in the era of exascale computing. In the laboratory, it is important to develop new techniques to measure energetic particles to permit quantitative investigations.

Due to the broad applications of magnetic reconnection, this Roadmap cannot possibly cover all aspects of the field.  In addition to the topics discussed here, there are other exciting research directions that we encourage readers to explore.  First, there are important connections with magnetic turbulence, another fundamental plasma process which has attracted a great deal of research.   In both MHD \cite{schekochihin20,zhdankin13,loureiro17,dong:2018,Boldyrev20} and kinetic regimes \cite{loureiro20,Boldyrev20,phan18}, the turbulent dynamics may lead to a plethora of current sheets that influence the turbulent cascade while also contributing to particle heating and acceleration.  Since plasma turbulence is ubiquitous throughout the Universe, this implies that magnetic reconnection may be active in a much wider range of applications, including collisionless shocks \cite{karimabadi14,matsumoto15} which often generate turbulence.   In this Roadmap, we primarily focused on low plasma $\beta (\lesssim1)$ regimes where available magnetic energy is comparable, or larger than, the initial thermal energy, and thus reconnection can power flares and accelerate particles to high energies. At higher $\beta$, however, the magnetic field can still play a critical role, as surveyed by in ref.\cite{ji:2011}. One particularly interesting example is that the saturation of the magneto-rotational instability due to reconnection~\cite{hawley92} can determine the level of angular momentum transport required to explain the observed fast accretion in astrophysical disks that drive high-energy jets. Lastly, magnetic reconnection under extreme conditions~\cite{uzdensky11} is not covered here either, but has recently attracted strong interest, motivated by recent discoveries such as fast radio bursts~\cite{Zhang:2020}. In order to explain these extreme events, various exotic effects need to be taken into account, such as ultra-relativistic effects and pair production in the strong fields environment, and general relativity near black holes.  However, the underlying multiscale physics discussed in this Roadmap is still expected to be important. This connects the reconnection events in the solar system and in the laboratory, where detailed studies are more feasible, to distant astrophysical reconnection across the Universe.  A list of major open physics questions of magnetic reconnection are listed in Box 2 with corresponding sections or references.

\section*{Multiscale nature of magnetic reconnection}

Like many other natural phenomena, magnetic reconnection involves multiple temporal and spatial scales, and interactions across these scales are fundamental. As described before, historically much attention was paid to one particular class of scales during a given stage of the reconnection research: MHD scales during the first stage and kinetic scales during the second stage. Now during the third stage, cross-scale interactions between global MHD scales to local dissipation scales (including kinetic scales) have been recognized as fundamental to understanding magnetic reconnection in space and astrophysical plasmas. In this section, we focus on the current understanding of the multiscale nature of magnetic reconnection and its future prospects. We divide our discussion into two parts: plasmoid-mediated coupling across multiple scales, including its realization in partially ionized plasmas, and the influence of magnetic turbulence on multiscale reconnection, as illustrated in the lower panels of Fig.\ref{Overview}.

\subsection*{Plasmoid-mediated multiscale reconnection \label{s:multi}}

The plasmoid instability~\cite{shibata:2001,loureiro07} of sufficiently elongated current sheets serves as a key proposed mechanism for multiscale reconnection in large systems, see the lower-left panel of Fig.~\ref{Overview}. Free magnetic energy builds up in these current sheets when the system is driven, often slowly, before it is released explosively during fast reconnection resulting in a global reconfiguration, sometime as a consequence of feedback between ideal evolution and reconnection. The plasmoid instability can trigger the onset of fast reconnection and sustain it steadily until a sufficiently large fraction of magnetic free energy is depleted. During the latter process, the various different reconnection regimes or phases, including those involving multiple scales required for large systems, are illustrated by the reconnection phase diagrams~\cite{daughton:2012,ji:2011,huang13,karimabadi13a,cassak13,le15,loureirouzdensky:2016,pucci:2017} in the parameter space of $S$ and normalized plasma size $\lambda \equiv L/\rho_s$, shown in Fig.~\ref{Phase}a. 

The phase diagram emerges naturally~\cite{ji:2011} from the evolution of different stages of reconnection research. The first stage focuses on MHD physics characterized solely by collisionality in terms of $S$, but valid only in the large $\lambda$ limit, whereas the second stage focuses on kinetic physics at small $\lambda$, but often in the collisionless limit ($S \rightarrow \infty$). Between these two limits in the parameter space where both values of $S$ and $\lambda$ are large (yet finite), different cross-scale coupling mechanisms have been proposed and are the subjects of the third stage research. The reconnection phase diagram captures the landscape of these regimes where most of the space and astrophysical reconnection occurs~\cite{ji:2011}.

We emphasize that the phase diagram here is meant to classify different fully nonlinear (quasi-)steady reconnection regimes which are closely related to, but distinct from, different reconnection onset mechanisms or different tearing instabilities of reconnecting current sheets. Thus, a recent theoretical development~\cite{bhat:2018} on the onset of a plasmoid instability in the semi-collisional regime, where $\rho_s$ is below $\Delta_{SP}$, but above the inner layer thickness of linear tearing instability, is shown as a dash line indicating the possible existence of multiple X-line reconnection for $S<S_c$. There are experiments~\cite{hare:2017} operating in this regime, but they are driven externally at large Alfv\'enic inflows rather than arising from the linear plasmoid instability.

The two traditional reconnection phases are single X-line collisional and collisionless ones, and the former is the well-known SP slow-reconnection model \cite{parker57a} whereas the latter is fast kinetic reconnection~\cite{birn01a}. They are separated by the condition $\Delta=\rho_s$ which translates~\cite{ji:2011} to $S=\lambda^2/4$. Both models have been successfully confirmed experimentally~\cite{ji98,ren05a,fox17} or observationally~\cite{mozer02a}. A continuous thermodynamic phase transition has been recently established~\cite{JaraAlmonte:2021b} from single X-line collisional to collisionless phases during anti-parallel reconnection at $E_{rec}=E_D\sqrt{m/M}$ over a wide range of electron-to-ion mass ratios, $m/M=1 \rightarrow 100$. Here $E_{rec}$ is reconnection electric field and $E_D$ is Dreicer electric field~\cite{dreicer59}. A new single X-line regime has been added in which reconnection is supported only by electron dynamics due to the small plasma sizes~\cite{phan18,Sharma2019}, which is important especially during collisionless plasma turbulence when energy cascades down to current sheets below ion scales~\cite{loureiro17,califano20}.

Most of natural and laboratory high-temperature fusion plasmas, however, fall into one of three multiple X-line phases~\cite{ji:2011}. Many of these plasmas are in the multiple X-line collisionless regime where kinetic effects play dominant roles throughout the reconnecting current sheet when the plasma size is sufficiently large~\cite{daughton06a}, $\lambda>\lambda_c \sim 50$. Familiar examples include Earth's magnetosphere and modern high-temperature tokamaks. Even for plasmas which are globally collisional (that is the mean free paths are short compared to the plasma sizes) locally in the reconnecting current sheet, particle dynamics can be essentially collisionless. This corresponds to the multiple X-line collisional-collisionless hybrid phases and its separation from the multiple X-line collisional phase, $S=(\sqrt{S_c}/2) \lambda$, is shown in Fig.~\ref{Phase}a for the case~\cite{ji:2011} where the number of plasmoids scales as $S/S_c$. Familiar examples of such plasmas include the solar corona, corona of accretion disks, and the magnetosphere of highly magnetized neutron stars (magnetars) which may host the recently observed fast radio bursts~\cite{Zhang:2020}. There is also indirect evidence of plasmoid formation in the multiple X-line collisional regime during the start-up of a spherical tokamak~\cite{ebrahimi15}.

The explosive release of even a portion of the magnetic energy can in principle heat the low-$\beta$ plasma substantially to power flares, and some of the charged particles can attain relativistic energies which are effectively collisionless, thus explaining the essential observational features.  Within weakly collisional regimes, the particle acceleration from reconnection often leads to a pronounced pressure anisotropy in the particle distribution with respect to the local magnetic field. The conditions for the generation of electron pressure anisotropy have been established in the single X-line regime~\cite{le15}, that is by demanding electron-ion collision time ($\tau_{ei}$) longer than reconnection time: $\tau_{ei} > \rho_s/(0.1 V_A)$ or equivalently $S>(5\beta M/m) \lambda$, where $0.1V_A$ is a typical reconnection inflow and $\rho_s$ is a typical local kinetic current sheet half thickness. This condition can be extended to the multiple X-line regime by replacing local reconnection time with its global counterpart $\tau_{ei} > L/V_A$ which leads to $S>(\beta/2)(M/m)\lambda^2$. Furthermore, since ions collide $\sqrt{M/m}$ times less frequent than electrons at comparable energies, similar conditions apply to ions.  Approximate threshold conditions for both electron and ion anisotropy are plotted in the updated phase diagram. Also shown are the corresponding conditions for electron runaway~\cite{ji:2011} when the reconnection electric field increases above $E_D$.

In order to test our understanding of multiple X-line regimes, new research efforts are needed including observations, simulations and laboratory experiments, each with sufficient resolution to validate fundamental ideas. The statistical properties of these phases are particularly important in determining the energetic consequences of magnetic reconnection, such as the nonthermal acceleration of electrons. Competing statistical models have been developed for power-law distributions of plasmoid size or flux~\cite{uzdensky10,fermo10,huang2012,loureiro12,TAKAMOTO13,guo13,LINGAM18,PETROPOLOU18,zhou2020,majeski21} with different power-law indexes.
Observationally, there is ample direct in-situ evidence for the existence of multiple X-line regime in Earth's magnetosphere~\cite{russell79,slavin03,chen:2008} and indirect evidence from remote-sensing solar observations~\cite{shibata95,mckenzie99}. However, the limited statistical studies of these observations as well as from the laboratory (see below) suggest exponential distributions~\cite{fermo11,guo13,dorfman13,VOGT14,olsen:2016,AKHAVAN-TAFTI18,bergstedt20}. It is unclear why this qualitative discrepancy exists, but multi-spacecraft missions with in-situ measurements, such as the current Magnetospheric MultiScale (MMS) mission~\cite{Burch2016} with many more satellites are required to better observe multiscale reconnection phenomena. Opportunities in laboratory experiments, observations, and computation are described in Box 3 and 4, respectively.

The upcoming FLARE (Facility for LAboratory Reconnection Experiments)~\cite{ji:2011} device represents a major opportunity in the next decade for the first laboratory accesses of the predicted multiple X-line phases with an extensive set of in-situ diagnostics. The first plasma operation~\cite{ji:2018} has already successfully demonstrated the feasibility of the design based on the existing Magnetic Reconnection eXperiment (MRX)~\cite{yamada97} in which much experimental work has been performed on single X-line reconnection phases~\cite{ji08,ren05a} (top panels of Fig.~\ref{Overview}). The parameter space expected to be accessible by FLARE is illustrated in Fig.~\ref{Phase}a, in comparison with that of the presently existing experiments including MRX and Terrestrial Reconnection EXperiment (TREX)~\cite{forest15}. These existing laboratory experiments mostly focused on single X-line regimes~\cite{yamada10,ji:2011} including a planned future upgrade on TREX to study electron pressure anisotropy at larger $S$. Although the generation of plasmoids in the reconnecting current sheet has been demonstrated in several experiments, most of these are in the collisionless regime with only a single plasmoid at a given time~\cite{stenzel86,ono11,dorfman13,olsen:2016,hare:2017} (top right pane of Fig.\ref{Overview}).  There is only one case where multiple plasmoids are observed, but on electron scale in the collisional regime~\cite{jaraalmonte:2016}. The FLARE experiment will enable dedicated explorations deep into the newly predicted multiple X-line phases. Experimental demonstrations of the transition into multiple X-line collisional and hybrid phases, as well as establishment of scaling properties in all three multiple X-line phases are of widespread importance.  By exploring this physics in the laboratory, we will be better positioned to extrapolate our understanding to astrophysical applications.

\subsection*{Multiscale reconnection in partially ionized plasmas\label{s:pi}}

As discussed before, the ultimate engine of reconnection is the ideal relaxation of stressed magnetic fields, which is enabled by non-ideal physics in localized diffusion regions. Ideal relaxation in the MHD description is well understood, but an interesting question is how does reconnection proceed when field-line relaxation itself is non-ideal? Globally, non-ideal processes introduce new characteristic length scales and physical processes that alter reconnection and couple to diffusion region physics in new ways. One rich, yet relatively unstudied, class of globally non-ideal plasmas is partially ionized systems.

In partially ionized systems, the ionization fraction, $\chi \equiv \rho_i/\rho$, is small, $\chi < 1$. Here the total density consists of ion and neutral densities, $\rho \equiv \rho_i+\rho_n$. This is the case in many space, astrophysical, and laboratory systems of interest, such as the solar chromosphere ($10^{-4} \lesssim \chi \leq 1$), molecular clouds ($\chi \lesssim 10^{-6}$), the interstellar medium ($10^{-4} \lesssim \chi \lesssim 10^{-2}$), and laboratory reconnection experiments ($\chi \gtrsim 10^{-2}$). Neutrals couple to the plasma through a variety of collisional processes including elastic scattering, excitation, deexcitation, radiation, ionization, and recombination~\cite{Ni2020}. The relative importance and detailed physics of each of these processes strongly depend on the plasma composition, however, ion-neutral elastic collisions are the most general and well-studied effect. 

Elastic collisions result in a frictional force, $\bm{F}_{coll}$ in Eq.~\eqref{ionMomentum}, which can dissipate intermediate-scale flows and MHD waves. Understanding how both the global reconnection process and the local diffusion region self-regulate in the presence of friction is a forefront question in partially ionized reconnection theory. Friction can be approximately included in single-fluid MHD theory via the ambipolar diffusion term~\cite{mestel56} which is equivalent to an enhancement of the perpendicular resistivity

\begin{equation}\label{eq:ambipolar_diffusion}
    \eta_{\perp,AD} = \mu_0 (1-\chi)\frac{V_{A^\star}^2}{\nu_{ni}}
\end{equation}

where $V_{A^\star} = B/\sqrt{\mu_0\rho} = V_A\sqrt{\chi}$ is the bulk Alfv\'en speed and $\nu_{ni}$ is the neutral-ion collision frequency. Since ambipolar diffusion does not produce a parallel electric field, it alone cannot change the magnetic topology or drive reconnection. Instead, ambipolar diffusion nonlinearly drives thin current sheets where reconnection may occur due to classical resistivity or kinetic effects~\cite{Brandenburg1994}. Thinning is ultimately halted~\cite{Brandenburg1995} by the effects from ion pressure or inertia at moderate $\beta$.  

Based on Eq.~\eqref{eq:ambipolar_diffusion}, the characteristic length scales for ambipolar diffusion can be defined via $\eta_{\perp,AD}/(\mu_0 V)$. In the small-$\chi$ limit and for $V=V_{A}$ and $V=V_A^\star$, these length scales are just $L_{in} \equiv V_A/\nu_{in}$ [$\nu_{in}=(\rho_n/\rho_i) \nu_{ni}$ is ion-neutral collision rate] and $L_{ni} \equiv V_{A^\star}/\nu_{ni}$ respectively. Single fluid theory is generally valid for the combined plasma-neutral system when $L > L_{ni}$ and for an uncoupled plasma, where neutrals can be neglected, when $L < L_{in}$. In the intermediate regime, $L_{in} < L < L_{ni}$, frictional dissipation is important and a multifluid approach is required. These collisional regimes are well known and, up to factors of order unity, apply to Alfv\'en waves~\cite{Kulsrud1969}, linear tearing modes~\cite{Zweibel1989}, SP current sheets~\cite{Zweibel1989,Malyshkin2011}, and multifluid reconnection \cite{Malyshkin2011}. Frictional dissipation is thought to be important during nonlinear reconnection in the multifluid regime~\cite{Malyshkin2011,Zweibel2011}, although this regime has not been extensively explored neither by numerical simulations nor by experiments.

Rather, the majority of numerical simulations have focused on the regime $L \gtrsim L_{ni}$ using either single-fluid\cite{Ni2007} or multifluid \cite{Smith2008,Leake2012,Murphy2015,Ni2018a,Singh2019} treatments. The inclusion of multifluid effects resulted in a modified SP scaling, $\Delta/L \sim (S^\star)^{-1.1}$ with $S^\star \equiv S\sqrt{\chi}$~\cite{Leake2012}. The plasmoid instability still occurred at $S^\star \geq 10^4$, and thus remains a leading candidate for bridging global and local scales in partially ionized plasmas. At smaller scales, the onset of fast, multifluid reconnection was predicted to occur when $\Delta \approx d^\star$ where $d^\star \equiv d_i/\sqrt{\chi}$ is the effective kinetic scale~\cite{Malyshkin2011}. However, multifluid simulations have not observed this transition \cite{Leake2013,Murphy2015,Ni2018a}.

Most existing experiments can only access the regime $L \lesssim L_{in}$ and have shown that fast, multifluid reconnection does occur \cite{Lawrence2013,Takahata2019}. The reconnection electric field was found to scale as $E \sim BV_{A^\star}$ such that neutrals remain important even at small scales~\cite{Lawrence2013}. Particle-in-cell (PIC) simulations with a kinetic treatment of both neutrals and collisional processes have shown a similar transition to fast reconnection when $\Delta \approx d_i \ll d^\star$~\cite{JaraAlmonte2019}. In both PIC simulations and experiments, the local current sheet where fast reconnection occurred had $L \sim L_{in}$. An attempt to reproduce these results with multifluid simulations instead found that the current sheet remained long, $L \sim L_{ni}$, and that the reconnection rate followed a SP scaling~\cite{jaraalmonte:2021}. Taken together, these results suggest that multifluid reconnection physics in partially ionized plasmas is not well understood. These regimes can be explored in the upcoming experimental facilities, such as FLARE, where all three partially ionized regimes are accessible.

Including partially ionized effects into the reconnection phase diagram requires introducing two additional physical parameters: $\chi$ and $L_{in}$ (since $L_{ni} \approx L_{in}/\sqrt{\chi}$). For clarity we consider a toy system: a 1D, semi-empirical model of the lower solar atmosphere~\cite{Avrett2008}. One dimensional models significantly oversimplify the rich dynamics present in the chromosphere~\cite{Carlsson2019}, but are useful pedagogical tools that give average values of $\chi$, $n_e$, and $T$ as a function of height, $h_{C7}$, above the model photosphere. Assuming $B_0 = 100$ Gauss, then the only free parameter in this model is the length of the current sheet, $L$. The reconnection phases can be mapped onto the 2D space of ($h_{C7}$, $L$) as shown in Fig. \ref{Phase}b. When $L < L_{in}$, the fully ionized phases \cite{ji:2011} are applicable. The regime $L_{in}(h) < L < L_{ni}$ maps to a `Multifluid' phase, but is not further subdivided. For $L > L_{ni}$, a new `Multiple X-line Hybrid Multifluid' is introduced, corresponding to a system which is globally plasmoid unstable ($S^\star \geq S_{crit}$), but contains embedded current sheets in the multifluid regime ($LS_{crit}/S^\star < L_{ni}$). Since the reconnection rate is determined by the smallest current sheets, the multifluid physics thus controls the reconnection rate in this regime. 

As shown by the regimes in Fig.~\ref{Phase}b, the solar chromosphere is both partially ionized and collisional. This is particularly important as it may be the only naturally occurring, collisional plasma where observations are detailed enough to test the plasmoid mediated reconnection theory. For example, it has been suggested that the shape of measured spectral lines could be due to plasmoids~\cite{innes15}. However, the plasmoid model is essentially 2D whereas the chromosphere has complex 3D structure~\cite{Carlsson2019}. 3D effects can introduce waves and instabilities (Box 1), or possibly allow for a fundamentally different turbulent reconnection mechanism~\cite{lazarian99,Lazarian2004}.

\subsection*{Influence of magnetic turbulence \label{s:turb}}

One of the most challenging uncertainties in the phase diagram concerns the influence of magnetic turbulence in large astrophysical systems, which can be either fully or partially ionized as discussed in previous sections.  Plasmoid mediated reconnection offers one possible scenario. However, even with laboratory validation, there will remain some outstanding questions regarding the general applicability in astrophysics.  First, both experiments and simulations are vastly smaller (in terms of $S$ or $\lambda$) than most astrophysical problems, and typically have too much symmetry (axi-symmetric, periodic boundary conditions).  Furthermore, astrophysical applications are expected to feature some pre-existing MHD turbulence, or neutral gas turbulence in partially ionized plasmas, which is difficult to include in experiments and is often neglected within simulations.   Finally, in large systems, such as the solar corona ($\sim 10^{8} d_i$), it is difficult to imagine that the global magnetic flux changes connectivity across a single kinetic scale layer as implied by the bottom left panel of Fig.\ref{Overview}.   Even within linear tearing theory, 3D dynamics (Box 1) allows the possibility of multiple resonance surfaces ($\bm{k} \cdot \bm{B}=0$) across a broader current layer \cite{galeev86}, thus permitting the global flux to change connectivity in a series of smaller steps, and perhaps leading to a `turbulent diffusion region' (TDR) (see lower right panel of Fig.\ref{Overview}) with thickness larger than kinetic scales ($\Delta \gg d_i,\rho_s$), but not necessarily large enough to permit a true MHD inertial range.   Assuming the mean outflow velocity remains $V_A$, mass conservation implies that the global reconnection rate is determined by the aspect ratio $\Delta/L$ of the TDR. Thus, it is crucial to understand the plasma physics that regulates this region. There are two limits to consider. First, the turbulence may arise spontaneously from the free-energy available within the large-scale magnetic shear, which is ultimately responsible for driving reconnection.  Alternatively, Alfv\'enic turbulence inherent within the global system might drive further turbulence within the diffusion region.  

Early work on the influence of magnetic turbulence was done within 2D MHD simulations \cite{matthaeus86} by injecting a spectrum of initial fluctuations, and more recently by driving fluctuations within high-$S$ simulations \cite{louriero09}.  The observed reconnection rate is enhanced significantly beyond the SP scaling in conjunction with the copious formation of plasmoids.  The most obvious explanation is that the fluctuations seed the plasmoid instability at higher level, thus reducing the critical Lundquist number and producing a faster rate $R\approx 1/\sqrt{S_c}$.  Although this simple interpretation is appealing, the physical relevance of these 2D simulations is not yet clear, since Alfv\'enic turbulence is inherently 3D and the plasmoid instability is also severely restricted in 2D models (for instance, magnetic islands (plasmoids) in 2D versus flux ropes in 3D; Box 1).

Within large-scale 3D simulations,  the evidence for spontaneously driven turbulence has accumulated in both MHD~\cite{oishi15,huang16,kowal17,beresnyak17,kowal20,yang20} and kinetic~\cite{daughton11a,leonardis13,daughton14,yhliu13a,nakamura16,dahlin17,li19,guo21} simulations and spacecraft observations \cite{Fu:2017}.  This turbulence is driven by tearing-type (plasmoid) instabilities, and further enhanced by the Kelvin-Helmholtz instability due to pre-existing flow-shear across the layer \cite{nakamura13} or arising from the reconnection outflow jets \cite{daughton14,pucci:2017b}.  In both fluid and kinetic simulations, this self-generated magnetic turbulence is dominated by coherent structures including flux-ropes, current sheets and flow vortices.   Unfortunately, the vast majority of these simulations use periodic boundary conditions, and the outflow jets quickly re-circulate leading to broader `turbulent boundary layers' with many interacting reconnection site and a thickness $(\Delta \gg d_i,\rho_s)$ that increases with time due to reconnection inflows.  These turbulent boundary layers are interesting and have potential applications\cite{nakamura:2017}, such as the lower-latitude boundary layer of the Earth's magnetopause.  However, the relevance to large-scale reconnection remains unclear since these simulations do not allow a macroscopic outflow jet to form.  Thus, these types of turbulent boundary layers do not represent a `turbulent diffusion region', which requires both inflow and outflow.     

Prior to the recent research on spontaneous plasmoid turbulence, researchers proposed that externally-driven  Alfv\'enic turbulence may play a crucial role in facilitating fast reconnection in large astrophysical systems \cite{lazarian99,lazarian20}. In contrast to plasmoids regulating the layer length $L$, this model assumes that turbulence broadens the layer thickness $\Delta$.  Furthermore, the model considers externally forced, isotropic turbulence which is weak at the injection length scale.  By using the Goldreich-Sridar theory for homogeneous MHD turbulence~\cite{goldreich97}, the reconnection rate is predicted to scale with the turbulent Alfv\'enic Mach number $R \propto M_A^2 = (V_{rms}/V_A)^2$ (with further dependence on injection length scale not shown).  This scaling has been confirmed within open-boundary 3D MHD simulations \cite{kowal12} in which turbulence is forced in a manner consistent with the above assumptions. Within these broader turbulent layers there is no obvious plasmoid dynamics.  However, more recent high-resolution MHD simulations \cite{yang20} have found a much weaker dependence of the rate $R \propto M_A$ on the injected turbulence, while also demonstrating ubiquitous plasmoid formation throughout the simulation.  One possible explanation is that the nature of the turbulence within strongly inhomogeneous reconnection layers is simply quite different from what was previously assumed ~\cite{lazarian99}.  Another possibility is that the injected turbulence seeds the plasmoid instability at a higher level, resulting in shorter current layers and faster reconnection, similar to previous 2D simulations with injected turbulence \cite{louriero09}.  With advances in computing, large-scale 3D kinetic simulations with injected turbulence have recently become possible~\cite{guo21}.  Even with injected turbulence, these simulations are dominated by flux ropes that repeatedly emerge within thin current sheets, leading to complicated 3D interactions including flux-rope kinking. In contrast to the MHD results, the observed reconnection rate $R \sim 0.1 $ appears to be independent of $M_A$.     

Although there is good evidence for turbulent boundary layers from both simulations and spacecraft observations \cite{nakamura:2017} of the lower-latitude boundary layer, there is presently no clear consensus regarding the influence of magnetic turbulence on large-scale reconnection layers.  However, with the emerging capabilities of exascale computers, we believe there are tremendous opportunities for progress.   Researchers must design their simulations with reliable open boundary conditions, that allow both Alfv\'enic inflow and outflow through the TDR.  To maintain the turbulence at a controlled level may likely require injection --- but preferably this should be done in the upstream region rather that directly within the TDR.  Finally, the simulations must be sufficiently large to permit the length and thickness of the TDR to form naturally, while also allowing a reasonable distance for the outflow jet.  Properly modeling the complex interplay between the TDR and the jet formation is critical for understanding turbulent reconnection.  At the present time, it is not clear whether thick TDRs can persist, or whether they spontaneously collapse back to thinner plasmoid dominated layers. Researchers must start with thicker turbulent layers to allow for both possibilities.  Another potential problem may be the formation of the outflow jet in the presence of strong turbulence that enhances momentum transport. In turbulence modeling, this enhanced transport is often represented by a `turbulent viscosity'.  Collisional viscosity $\nu$ reduces both the outflow and the reconnection rate by factor $(1+\nu \mu_0/\rho \eta)^{-1/2}$ that depends on the ratio of viscous to resistive dissipation (magnetic Prandtl number)~\cite{park1984}.  By analogy, one might expect the turbulent magnetic Prandtl number, $P_m$, to play a critical role in the jet formation.   If the turbulence leads to regimes where $P_m\gg1$ it may become difficult to form an outflow jet.  By properly designing the problem setup, there are ample opportunities to make progress on answering all of these questions. In the laboratory, there is need to develop reliable techniques to induce turbulence at desirable places in the plasma (and also in neutral gas, when partially ionized) to test all these ideas. 

\section*{Onset of explosive reconnection events \label{s:onset}}

Reconnection events in natural and laboratory plasmas (Fig.~\ref{Example}) are often explosive, featuring a sudden transition from a phase of slow magnetic energy build-up to a phase of rapid energy release.  On the Sun, for example, turbulent motions of field lines at the solar photosphere can cause magnetic energy to build up slowly in the corona~\cite{kusano:2002} in the form of twisted filaments, which can be stable for long periods before their abrupt and catastrophic loss of equilibrium. Solar flares that are associated with these filament eruptions are characterized by a rapid impulsive phase of energy release on the order of $\sim 100$ seconds. In the Earth's magnetosphere, the explosive behavior is also observed in substorm onset~\cite{sitnov:2019}. Reconnection at the dayside magnetopause can lead to magnetic flux accumulation in the lobes of the magnetotail plasma sheet, where the energy can build up gradually and be sporadically released via tail reconnection~\cite{angelopoulos:2008}. These distinct phases are also observed in laboratory plasma environments. Sawtooth oscillations in tokamaks~\cite{vongoeler:1974,hastie:1997} and reversed field pinch~\cite{prager05} in magnetically confined plasmas are characterized by repeated cycles of slow build-up of core pressure and magnetic shear on slow transport time-scales, followed by a rapid crashes in which these profiles are relaxed. A full understanding of these explosive onset behaviors needs to address the role of reconnection in the magnetic energy build up phase and the fast energy release phase, as well as the `trigger mechanism' that causes the transition. These will necessarily differ, depending on the application, but much can be learned about the basic reconnection physics by studying highly simplified, yet representative onset scenarios.

\subsection*{Disruption of a thinning current layer by tearing instabilities\label{tearingonset}}

In one scenario, shown Fig.~\ref{onsetfig}a, free magnetic energy builds up in a current sheet that forms and thins dynamically in response to some external drive. The thinning sheet passes a stability boundary at which tearing-type (reconnecting) instabilities become unstable, and these trigger fast reconnection onset when they grow large enough to nonlinearly disrupt the current layer. Examples include collisionless tearing~\cite{pritchett:1995} in the magnetotail current sheet, magnetic islands produced in driven laboratory reconnecting layers~\cite{ono11,dorfman13}, and the disruption of dynamically thinning MHD-scale current layers by the plasmoid instability.

Following results that the plasmoid instability is super-Alfv\'enic~\cite{loureiro07}, $\gamma \tau_{A} \sim S^{1/4}$ where $\tau_A\equiv L/V_A$, for current sheets with the SP aspect ratio, $L/\Delta = S^{1/2}$, it was quickly realized that dynamically thinning current layers at high $S$ may disrupt before the SP sheet can be formed. The foremost questions then concern how large the aspect ratio can be, and how many plasmoids disrupt the layer. To address these questions, theoretical models have been proposed~\cite{puccivelli:2014,uzdenskyloureiro:2016,comisso:2016,huang:2017} that apply results from tearing instability theory to dynamically thinning current sheets with variable aspect ratio $L/\Delta$, where the tearing growth rates increase as the layer thins. Initial estimates assumed that disruption would occur when the growth rates $\gamma \tau_{A} \approx 1$, finding a critical aspect ratio of $L/\Delta \approx S^{1/3}$ for the resistive MHD model~\cite{puccivelli:2014}. Subsequent models have shown the importance of the level of background noise that can seed the instability, the specific time-dependent profiles for the current layer thinning~\cite{comisso:2016}, and the stabilizing role by the reconnection outflow jets~\cite{ni:2010,huang:2017} in setting the threshold between single and multiple X-line reconnection in the phase diagram (the green horizontal line in Fig.~\ref{Phase}a). Taking into account these effects over the full time history of the growing modes can be done with a principle of least time approach~\cite{comisso:2016,huang:2017}, which yields the dominant mode at disruption.

In the coming decade, new well-diagnosed laboratory experiments such as FLARE (Fig.~\ref{onsetfig}a) may be able to probe this physics in unprecedented detail. Towards this goal, it is important to extend the theories to include additional physical effects that are present in laboratory experiments and first-principles (kinetic) simulations. Studies of fixed aspect ratio SP layers find faster plasmoid growth rates when the inner tearing layer thickness is ordered as $\delta_{in} \ll (d_i,\rho_s)$ for high-$\beta$ ~\cite{baalrud:2011} and low-$\beta$~\cite{bhat:2018} regimes, respectively. Simulations have shown these effects can introduce a lower threshold for multiple X-line reconnection (semi-collisional threshold in Fig.~\ref{Phase}a) that may explain plasmoids observed in kinetic simulations~\cite{daughton09b} and laboratory experiments~\cite{olsen:2016,jaraalmonte:2016,hare:2017} at lower Lundquist numbers than predicted by resistive-MHD alone. Successful validation of onset models against experimental data will also likely require the consideration of complex collisional transport effects, such as neutral-plasma collisions, temperature-dependent transport coefficients, and Dreicer runaway~\cite{daughton09a,stanier:2019}.  Including 3D effects in these thinning current sheet models is also critical to model real systems (Box 1). For example, 3D configurations can permit a spectrum of tearing modes at different oblique angles~\cite{daughton11a,baalrud:2012}, which, in some cases, can have faster growth rates than the usual 2D modes. Depending on the specific current sheet plasma profiles, the rational surfaces for these oblique modes can have significant density or temperature gradients, and may be subject to diamagnetic stabilization~\cite{baalrud:2018,stanier:2019}. Understanding how these complex effects combine, or compete, in dynamically thinning current sheets and lead to the onset of fast reconnection is a challenging task. Some initial studies have been conducted in this area~\cite{pucci:2017}, but there is still much to be learned. 

Despite its broad applicability, one limitation in applying this scenario to realistic systems --- such as those in Fig.~\ref{Example} --- is that it only includes the reconnecting current sheet and neglects the details of the larger scale system that supplies the magnetic flux. In particular, there may be two-way feedback between the reconnection layer and the larger scale system that can play a key role in the evolution of the reconnecting system.

\subsection*{Coupling of reconnection to macroscale instabilities}

One scenario which includes this two-way feedback between the reconnecting current sheet and the macroscale system is shown in Fig.~\ref{onsetfig}b. In this scenario, the linear instability that triggers the explosive event may be predominantly of ideal nature. The free energy source driving it can also be magnetic, such as the coalescence instability (Fig.~\ref{onsetfig}) or the kink instability, or may be from another source such as a thermal pressure gradient. Reconnecting current sheets that form in the nonlinear phase of this instability still play a critical role in the energy release, and allow the system to relax to a state of much lower free magnetic energy at saturation. It is important to note that this scenario can be complementary to the one discussed in the previous subsection, in that tearing-type instabilities can occur in the current sheet that forms and thins during the nonlinear phase of the macroscale instability.

\begin{figure*}[t]
\centering{\includegraphics[width=0.8\textwidth]{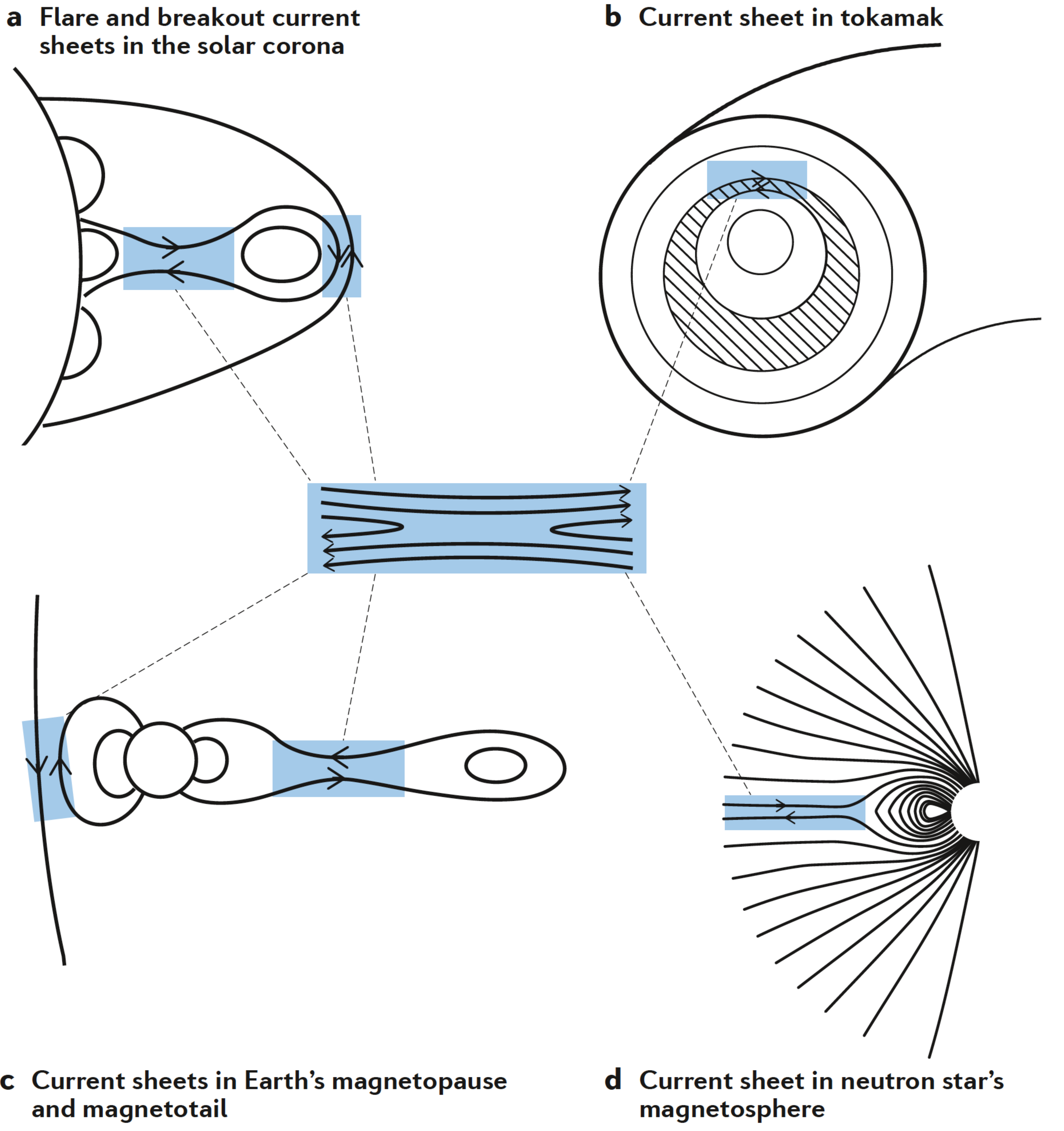}} 
\caption{
Examples of large-scale, electric current sheets in space, solar, astrophysics and laboratory fusion plasmas. a. On the Sun, convection in the photosphere moves and twists field lines, producing magnetic shear in the corona. The loss of equilibrium and eruption of a large flux-rope can drive reconnection at an underlying flare current sheet~\cite{shibata95}. Some flares may also have an overlying break-out current sheet where reconnection can remove the stabilizing effect from the stretched overlying magnetic field-lines. b. In laboratory Tokamak fusion experiments, instabilities can rapidly deform the internal plasma and create current sheets~\cite{vongoeler:1974}. c. The solar wind carries an interplanetary magnetic field which can form current sheets at the Earth's magnetopause and within its magnetotail~\cite{dungey61}. d. In compact astrophysical objects, such as neutron stars, rapid rotation can generate extended current sheets outside of the light cylinder in their magnetospheres~\cite{spitkosky06}. 
\label{Example}
}
\end{figure*}

\begin{figure*}[t]
\centering{\includegraphics[width=1.0\textwidth]{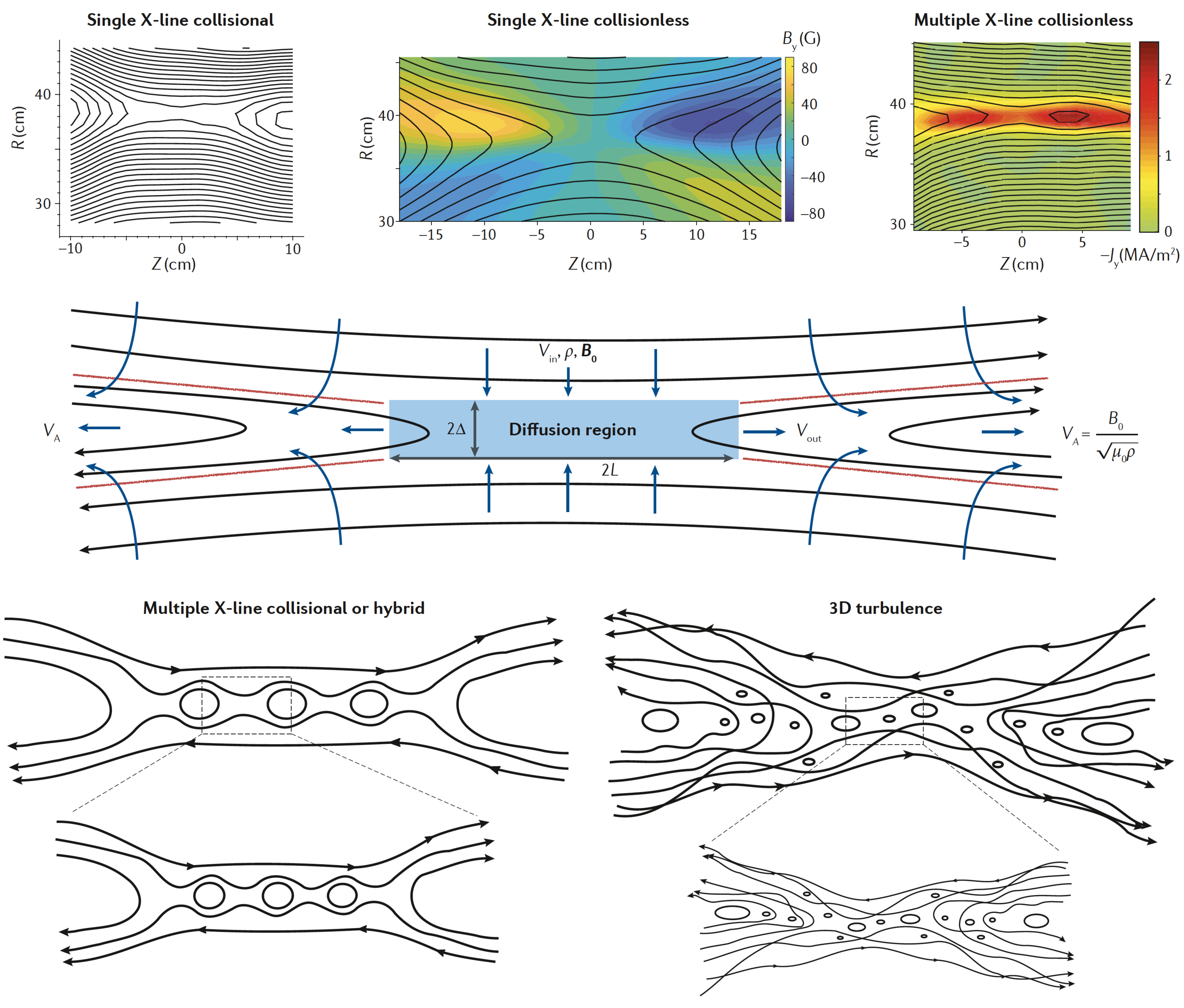}} 
\caption{Magnetic reconnection within a current sheet (depicted at the center) features inflows ($V_{in}$) of plasma with density $\rho$ and reconnecting magnetic field $B_0$, outflows ($V_{out}$) on order of Alfv\'en speed $V_A \equiv B_0/\sqrt{\mu_0 \rho}$ where $\mu_0$ is vacuum permeability, and a diffusion region (with a length of $2L$ and a thickness of $2\Delta$) where field lines change connectivity. Magnetic separatrices (red) are field lines marking the topological boundary between the upstream flux and the downstream flow jet.  Top panels display reconnection phases that have been confirmed in the laboratory experiments: single X-line collisional~\cite{ji98}, single X-line collisionless~\cite{yoo13}, and multiple X-line collisionless~\cite{dorfman13}. Here $R$ and $Z$ are radial and axial coordinates, respectively, in Magnetic Reconnection Experiment (MRX)~\cite{yamada97} where these data were taken. $B_Y$ and $J_Y$ are out-of-the-plane field and current density, respectively. Bottom panels illustrate cartoons of plasmoid mediated and 3D turbulent reconnection, both of which are expected to exhibit some degree of self-similarity. The top left panel is reproduced with permission from Ref. \cite{ji98}, the top right panel is reproduced with permission from Ref. \cite{dorfman13}. \label{Overview}}

\end{figure*}

\begin{figure*}[t]
\centering{\includegraphics[width=1.0\textwidth]{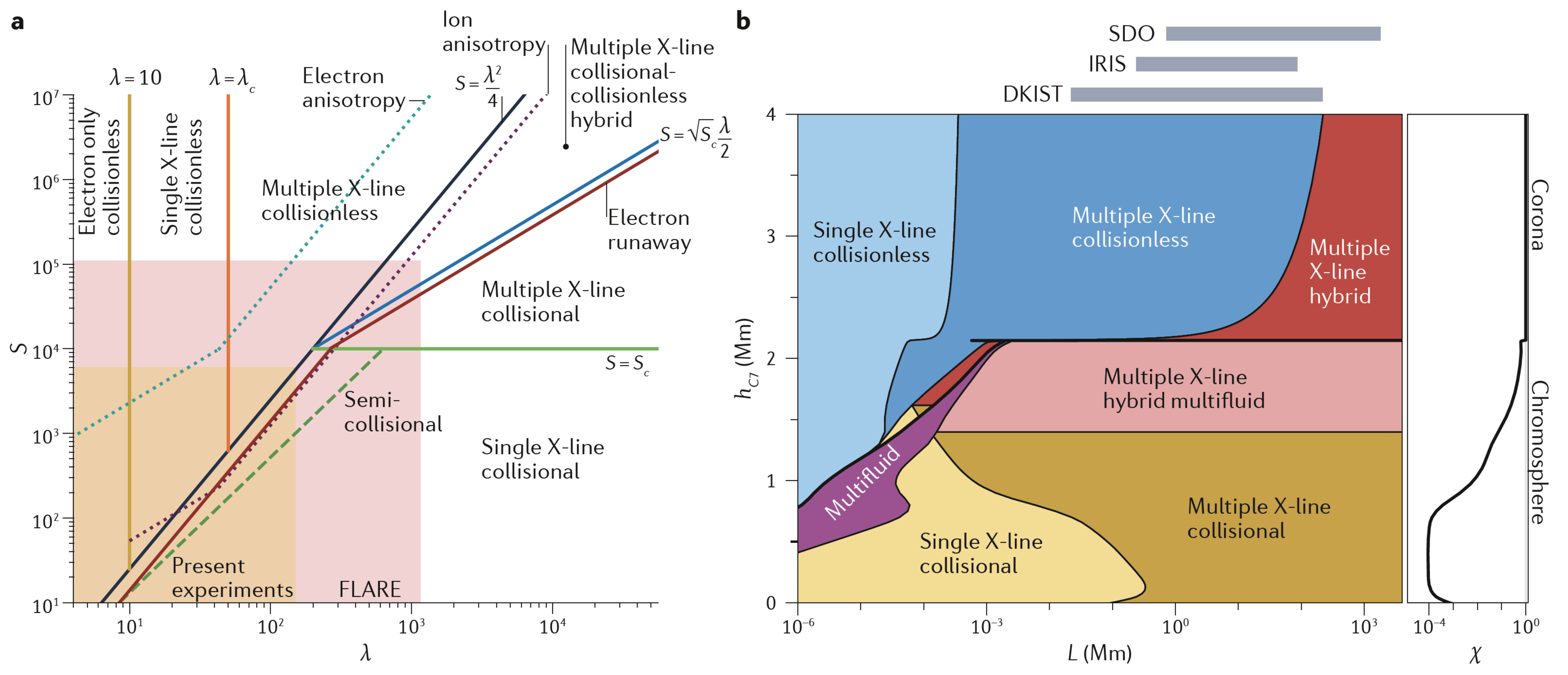}} 
\caption{Phase diagram of magnetic reconnection. a. Phase diagram in the parameter space of normalized plasma size $\lambda$ and Lundquist number $S$. In addition to two traditional single X-line phases (collisional and collisionless), there are three multiple X-line phases based on plasmoid instability of reconnecting current sheet when either or both of $\lambda$ and $S$ are sufficiently large. These three multiple X-line phases are classified based on the plasma collisionality on relevant scales, either collisionless or collisional across all scales, or collisional on larger scales, but collisionless on smaller scales within the same current sheet (termed hybrid). $S_c$ is the critical Lundquist number above which MHD current sheet is unstable to form plasmoids. Other than these 5 phases, a regime in which reconnection occurs only by electron dynamics due to small plasma sizes, as well as regions in which electron and ion pressure anisotropy could be dynamically important during reconnection time, are added. Also shown are parameter space already demonstrated by presently existing laboratory experiments and parameter space expected to be accessible by the upcoming experiment, FLARE, for detailed experimental explorations of the newly predicted multiple X-line phases directly relevant to heliophysical, astrophysical, and laboratory fusion plasmas.
b. An example application of the reconnection phase diagram to the 1D, semi-empirical C7 model of the lower solar atmosphere~\cite{Avrett2008} assuming a reconnecting field strength of 100 G. $h_{C7}$ is the height from Sun’s photosphere according to the C7 model. The condition $L = L_{in}\equiv V_A/\nu_{in}$ ($\nu_{in}$ is the ion-neutral collision rate) is shown with a thick black line and separates the regimes of decoupled (upper half) and strongly coupled (lower half) plasma and neutral gas dynamics. Upper grey bars show the resolution and field of view for example ground based telescopes (DKIST, Daniel K. Inouye Solar Telescope) and satellites (SDO, Solar Dynamics Observatory and IRIS, Interface Region Imaging Spectrograph). Right panel shows the average ionization fraction $\chi \equiv \rho_i/(\rho_i + \rho_n)$.
\label{Phase}}
\end{figure*}

\begin{figure*}[t]
\centering{\includegraphics[width=0.8\textwidth]{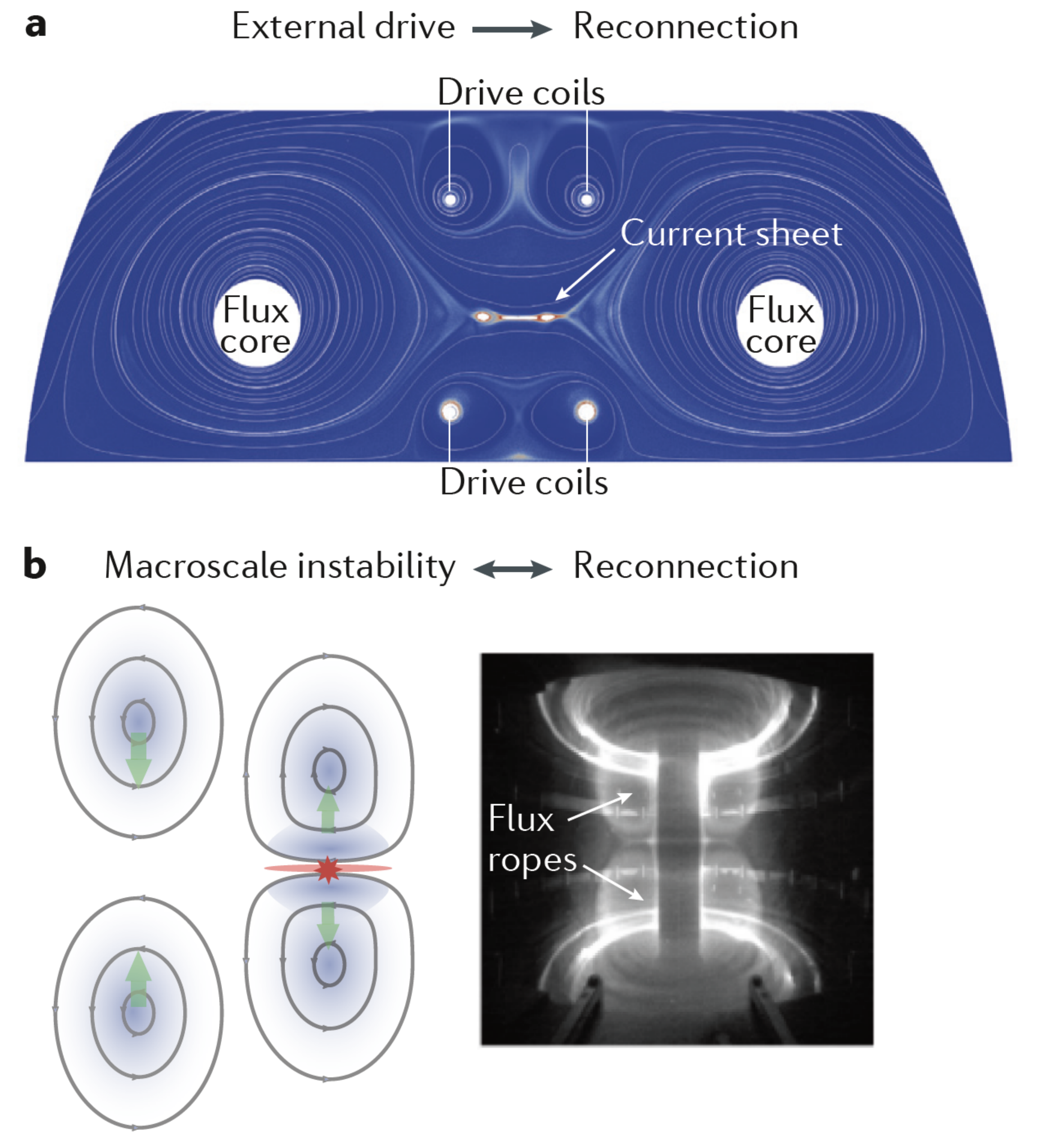}} 
\caption{Two reconnection scenarios with different onset/trigger mechanisms and macroscale coupling. The direction of the red arrows indicates the direction of coupling/feedback. a. Numerical simulation of driven reconnection in the FLARE laboratory experiment. Ideal build-up of magnetic energy in a current sheet occurs due to inductive current drive from the drive coils and the flux cores. b. The coalescence of magnetic islands (flux-ropes in 3D). On the left: the macroscale islands collide due to their parallel currents (blue), forming a reconnecting current sheet (red). The self driven process can cause significant magnetic pile-up in regions upstream of the current sheet that can cause islands to bounce off each other and undergo sloshing oscillations. On the right: fast camera image of the coalescence of two toroidal flux-ropes used for `merging-compression' start-up in the Mega Ampere Spherical Tokamak (MAST) experiment. The final coalesced state forms a spherical tokamak after two flux ropes merge by reconnection~\cite{tanabe2017}. The right side of panel b is reproduced with permission from the UK Atomic Energy Authority. \label{onsetfig}}
\end{figure*}

Examples of this scenario in space include the ballooning instability in the magnetotail current sheet leading to localized thinning and reconnection~\cite{pritchettcoroniti:2013}, as well as Kelvin-Helmholtz driven reconnection at the magnetopause flanks~\cite{eriksson:2016, nakamura:2017}. In the standard model of eruptive solar flares~\cite{priest00,shibata:2011}, the bulk of the free magnetic energy is stored in the form of a twisted filament. The eruption may be triggered by a loss of equilibrium~\cite{forbes:1995} or by macroscale kink~\cite{hood:1979} or torus~\cite{kliem06} instabilities. The formation and subsequent reconnection at an underlying flare current sheet (Fig.~\ref{Example}a without the break-out reconnecting current sheet) can significantly accelerate the erupting filament.
Finally, in laboratory magnetic fusion experiments, models of the sawtooth oscillation typically involve driven reconnection of helical flux at a current sheet formed in response to the internal kink instability~\cite{kadomtsev75a,porcelli:1996}, although other explanations have been proposed~\cite{jardin:2020}.

A critical unresolved issue for the present scenario is to understand the potentially two-way feedback between the macroscale instability and the diffusion region microphysics at the reconnecting current sheets. An illustrative example is the island coalescence instability (Fig.~\ref{onsetfig}), which can show distinct behavioral regimes dependant on this two-way feedback. This has been studied with a variety of plasma modeling tools. In resistive MHD, three regimes are possible depending on $S$: the islands can merge with a rate independent of dissipation~\cite{biskamp:1980} for $S\leq 10^3$, exhibit `sloshing' for $10^{3}\lesssim S \lesssim 10^{5}$ where reconnection repeatedly switches on and off~\cite{biskamp:1980,knoll:2006a}, and have a plasmoid unstable diffusion region for $S \gtrsim 10^5$. In this latter regime, the `peak' reconnection rate becomes fast~\cite{shadid:2010}, but the total coalescence time and the presence of sloshing is not yet well understood. Including two-fluid effects can reduce pile-up and sloshing in small systems~\cite{dorelli:2003, knoll:2006b}, but extended-MHD, hybrid, and kinetic studies of larger systems have shown sloshing can occur due to the magnetic flux pile-up outside of the ion diffusion region~\cite{karimabadi:2011,stanier:2015,ng:2015,makwana:2018}. When guide field is weak compared with reconnecting field, the behavior of the system is strongly dependent on the ion kinetic physics~\cite{stanier:2015,ng:2015}, which needs to be retained on global scales in order to reproduce the correct dynamics~\cite{makwana:2018}.

The coupling between reconnecting current sheets and ideal MHD instabilities can be more complex than the island coalescence problem described above, and understanding this physics may be critical to a number of puzzling phenomena in reconnecting systems. Examples include the long-standing problem of apparently incomplete reconnection in tokamak sawtooth oscillations~\cite{hastie:1997}, with one possible explanation being that the formation of thin current sheets and reconnection may enable secondary ideal instabilities such as ballooning (or interchange)~\cite{chapman2010} that may interrupt the complete reconnection of the helical flux, or may even lead to disruptions. Another puzzling phenomenon concerns the observed `failed eruptions' both on the Sun~\cite{sun15} and in the lab~\cite{myers15} possibly due to the nonlinear feedback from reconnection internal to the flux ropes, which are initially unstable due to the torus instability. 

Finally, complexity may arise from the coupling of multiple reconnection sites with the macroscale instability. An example of this is the `break-out' model~\cite{antiochos:1999} shown in Fig.~\ref{Example}a, which has had success in explaining some features of eruptive solar flares and frequent smaller-scale solar jets~\cite{wyper:2017}. In addition to the standard model of eruptive flares described above, reconnection can also occur at a break-out current sheet above the emerging filament. This can remove the stabilizing magnetic tension due to strapping magnetic field-lines, to trigger a catastrophic loss of equilibrium~\cite{longcope:2014}. As the filament erupts, it can drive reconnection at the underlying flare current sheet as in the standard model. High resolution studies with solar observations~\cite{kumar:2019} and MHD simulations~\cite{karpen:2012,wyper:2017} have found large numbers of plasmoids or blobs in both of these break-out and flare current sheets, but the precise role of plasmoid-mediated fast reconnection on the timing and energetics of each stage of the eruption are not yet fully understood.

These types of problems have proved to be particularly challenging to understand in the past due to the multiscale nature of this coupling between the macroscale instability and the reconnecting current sheets. There is a real opportunity to make progress in the next decade, using advances in multiscale techniques in experiment and observation (Box 3), as well as in simulation (Box 4). Also, progress can continue to be made from the study of idealized systems that are highly simplified, yet retain some of the important physics by using properly designed numerical and laboratory experimental settings.

\section*{Particle acceleration and heating \label{s:accel}}

One important consequence of magnetic reconnection is that it rapidly converts stored magnetic energy into particle energy in the forms of flows, bulk (`thermal') heating, and high-energy (`nonthermal') tails. There are laboratory measurements of energetic particle generation during reconnection events in magnetically confined plasmas. In particular, nonthermal electrons have been observed during magnetic reconnection from the sawtooth oscillation in a tokamak \cite{savrukhin:2001}.  In the Madison Symmetric Torus experiment, magnetic reconnection produces an anisotropic nonthermal electron tail population \cite{dubois:2017} as well as anisotropic ion heating and an energetic ion tail \cite{magee:2011}. Meanwhile, remote X-ray observations of solar flares, show that up to a half of the converted magnetic energy is released into nonthermal particles \cite{emslie:2004,krucker2010}. There have also been numerous in-situ observations of energetic particles associated with magnetic reconnection in planetary magnetosphere, including observations with electron power-law distributions up to 300 keV \cite{oieroset:2002}. Extrapolating to more extreme astrophysical environments, reconnection possibly including plasmoids is a promising candidate for accelerating some of the highest-energy particles observed in the Universe, such as those that radiate from pulsars and astrophysical jets \cite{cerutti:2012,sironi:2014,philippov:2019}. Other magnetospheric observations, on the other hand, show relatively limited bulk heating \cite{phan:2013} with a negligible contribution from nonthermal particles. Better understanding such differences based on the system parameters and size, particularly for electron energization, is an important goal in reconnection research.

Recent reviews describe the current understanding of electron acceleration during magnetic reconnection in the non-relativistic \cite{dahlin:2020,li2021} and relativistic regimes \cite{guo:2020}. First-principles numerical simulations suggest power-law nonthermal particle populations are more prevalent in systems with low $\beta$ and moderate guide magnetic fields, with acceleration efficiency enhanced in 3D compared to simplified 2D calculations. Multiple acceleration mechanisms have been studied. In the localized diffusion region near an X-line without a guide field, particles may become demagnetized, prone to be directly accelerated by the reconnection electric field as they undergo so-called Speiser or meandering orbits \cite{zenitani:2001,uzdensky11d}. There is recent laboratory evidence for electron acceleration by this mechanism in action during magnetically driven reconnection at low-$\beta$ using laser-powered capacitor coils~\cite{chien2022}. There is also numerical evidence~\cite{zhang21} that a substantial fraction of dissipated magnetic energy is carried by particles accelerated by this mechanism when they are demagnetized and escape from plasmoids in 3D.

\begin{figure*}[t]
\centering{\includegraphics[width=1.0\textwidth]{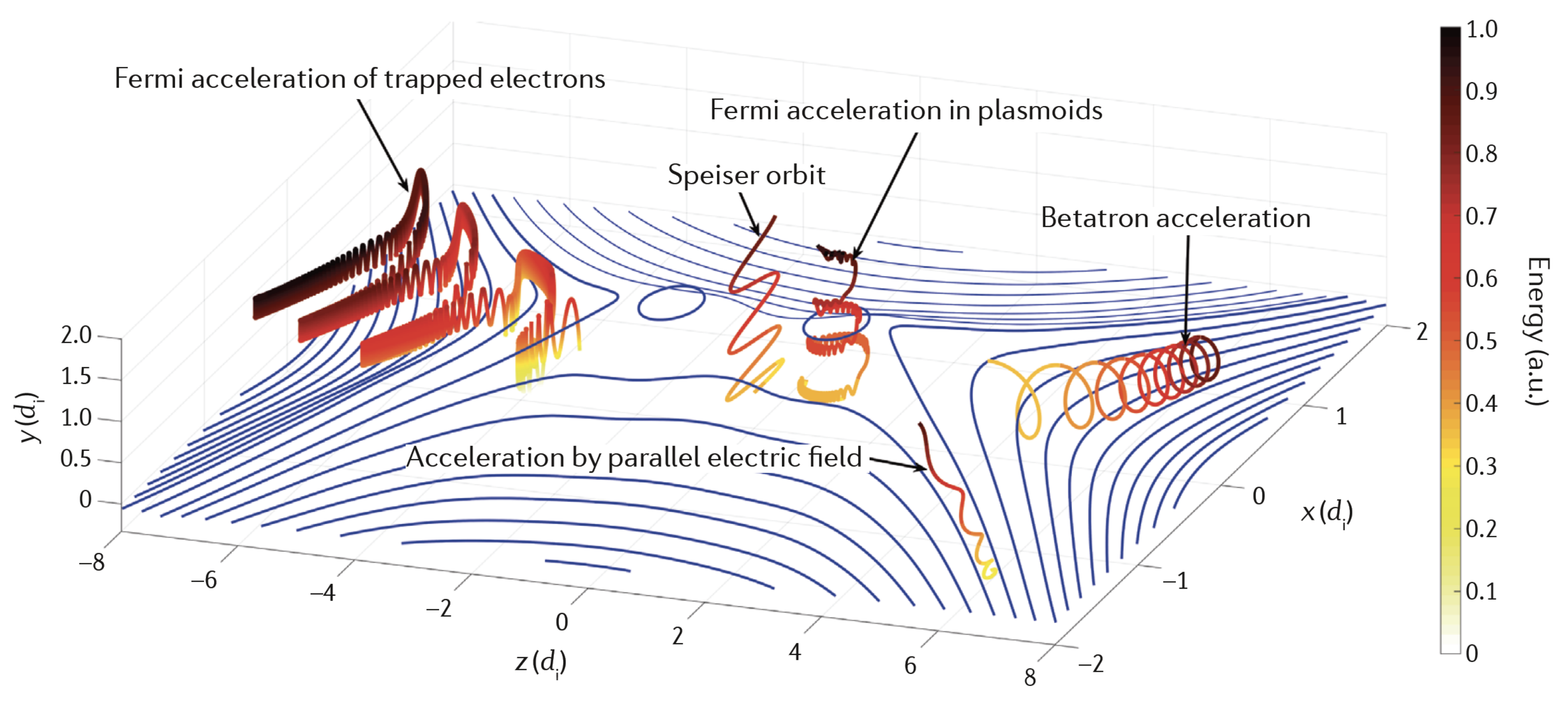}} 
\caption{Illustration of particle acceleration mechanisms. The reconnection electric field is along the $y$ direction. Acceleration by the reconnection electric field occurs near the X-line, which is associated with the Speiser orbit, and by parallel electric field near separatrices, which are boundaries between upstream (before-reconnection) and downstream (after-reconnection) regions. When particles near the X-line interact with the parallel electric field, ions (electrons) are accelerated along the (opposite) direction of the field, and the direct acceleration term $qE_\|v_\|$ (the first term on the right side of Eq.~\eqref{eqn:guide}) is positive. The betatron acceleration mechanism happens when particles flow into the stronger magnetic field regions in the exhaust. In this example trajectory, particles move along the positive $z$ direction by the $\bm{E}\times\bm{B}$ motion and $\nabla B$ is also along the positive $z$ direction. As a result, the second term of Eq.~\eqref{eqn:guide} is positive. Fermi acceleration occurs in the regions with high-curvature magnetic field lines. In this example, plasmoids are elongated along the $z$ direction, such that the field lines have the highest curvature at $x=0$. Inside the plasmoid, particles obtain a large velocity kick along the $y$ direction due to a large curvature drift and obtain a significant energy from the reconnection electric field [$q\bm{E}\cdot\bm{u}_c>0$ in Eq.~\eqref{eqn:guide}]. In the exhaust region, trapped particles gain energy whenever they pass the high curvature regions at $x=0$. $d_i$ is ion skin depth. }
\label{fig:heating}
\end{figure*}

Over larger scales when the particles remain magnetized, there are three basic particle acceleration processes that dominate in collisionless reconnection: Fermi acceleration~\cite{drake06a,dahlin14,guo14} which takes place as particles stream along and drift in relaxing curved magnetic field lines, relatively localized electric fields $E_\|$ parallel to the magnetic field directly accelerate particles~\cite{egedal:2013}, and betatron heating~\cite{hoshino01} which occurs as particles drift into regions of stronger magnetic field while conserving the first adiabatic moment $\mu = mv_\perp^2/B$. 
These mechanisms are described by guiding center equations of motion for the particles, and the energization rate in the non-relativistic limit is given approximately by the following equation \cite{northrop1963}: 
\begin{equation}
\label{eqn:guide}
    \frac{d\epsilon}{dt} = qE_\|v_\| +
    {\mu}\frac{dB}{dt} +
    q\bm{E}\cdot\bm{u}_c + 
    \frac{1}{2}m\frac{d}{dt}|\bm{u}_E|^2,
\end{equation}
where $\epsilon$ is the energy of a particle, $\bm{u}_E$ is the $\bm{E}\times\bm{B}$ drift velocity, ${d}/{dt} = {\partial}/{\partial t} + \bm{u}_E\cdot\nabla$, $E_\|$ is the parallel electric field, and $v_\|$ is the drift-corrected guiding center parallel velocity. For slowly varying fields, $\bm{u}_c$ is the curvature drift of particles and reduces to $\bm{u}_c\sim (mv_\|^2/qB)(\bm{b}\times\bm{\kappa}$), where $\bm{\kappa} = \bm{b}\cdot\nabla\bm{b}$ is the magnetic field curvature and $\bm{b}$ is the unit vector in the direction of the magnetic field. The terms on the right-hand side represent energy gain by the parallel electric field, betatron heating, Fermi acceleration, and the polarization drift respectively. The first three mechanisms are sketched in Fig.~\ref{fig:heating}. The polarization drift is particularly important for lower-energy ions \cite{li:2019}, and this term accounts for the kinetic energy gain in the large-scale reconnection outflow jets. Although Eq.~\eqref{eqn:guide} is valid in the non-relativistic limit, the same basic acceleration mechanisms are responsible for energizing particles in more extreme relativistic contexts relevant to astrophysical systems \cite{guo15,guo:2020,uzdensky11}.

Particles gain energy when their drift induced by the curvature of the magnetic field lines is aligned with the reconnection electric field~\cite{drake06a,guo14,dahlin:2020}, which is a form of Fermi heating. Fermi acceleration may resolve two major challenges for heating models. First, it may operate over large volumes in the presence of many magnetic islands or plasmoids \cite{shibata:2001}, which may be sampled by electrons in the presence of 3D ergodic field lines \cite{dahlin:2020}. There are several space observations of energetic electrons near magnetic islands or plasmoids \cite{chen:2008,zhong:2020}. Second, the Fermi process naturally generates power laws because a particle undergoing Fermi acceleration gains energy at a rate proportional to its energy, $\Delta \epsilon \propto \epsilon$. In the non-relativistic limit, adiabatic particle energization by Fermi acceleration along with the other heating mechanisms for magnetized particles leads to a Parker-like \cite{parker:1965} transport equation for the distribution $f(v_\|,v_\perp)$ (velocity directions are with respect to the magnetic field) of particles within a magnetic flux tube including time-varying plasma density $n$ and magnetic field $B$: \cite{montag:2017,drake06a}:
\begin{equation}
\frac{\partial f}{\partial t} + \frac{\dot{B}}{B}v_\perp^2\frac{\partial f}{\partial v_\perp^2} + \left(\frac{\dot{n}}{n} - \frac{\dot{B}}{B} \right)v_\|\frac{\partial f}{\partial v_\|} = \frac{D\{f\}}{\tau_{diff}} - \frac{f}{\tau_{esc}} + \frac{f_{inj}}{\tau_{inj}}
\end{equation}
where $D\{f\}/\tau_{diff}$ includes diffusive processes such as pitch-angle or energy scattering, $\tau_{esc}$ is a typical time for particles to escape the reconnection geometry, and $f_{inj}/\tau_{inj}$ is a source of newly injected particles. This transport equation leads to power-law solutions under certain assumptions, though the Fermi process becomes much less effective at generating power-law nonthermal tails in the presence of a significant guide (perpendicular to the reconnection plane) magnetic field component \cite{montag:2017}. With certain assumptions, it is possible to solve the Parker equation using the velocity and magnetic fields from MHD simulations to obtain predictions of the particle acceleration~\cite{li:2018}, but this approach does not include the self-consistent feedback.

Originally, direct acceleration by a parallel electric field $E_\|$, was considered only for particles within the diffusion region itself.  For example, there is an early laboratory observation of high-energy tail electrons \cite{stenzel86} generated in this way. At first glance, direct acceleration by $E_\|$ appears unimportant for large systems because only a very small fraction of electrons pass through the microscopic diffusion region. Nevertheless, differences between electron and ion dynamics may generate larger-scale fields $E_\|$ in boundary layers or separatrices~\cite{lapenta15} that extend 100s of kinetic scales \cite{egedal:2013}. The cumulative effect of $E_\|$ is measured by an effective potential defined as $\phi_\| = \int E_\| dl$, where the integral is along magnetic field lines. Theory and simulation show that the maximum of this potential scales as $e\phi_\|/T_e\propto 1/\sqrt{\beta}$ under laminar conditions, as observed in simulations and inferred from spacecraft data \cite{egedal:2013}. Besides acting on its own, direct parallel acceleration takes part in an interplay with the Fermi heating process.  First, the effective potential $\phi_\|$ aids Fermi acceleration by confining heated electrons within the reconnection exhaust \cite{egedal:2015} to allow them to undergo multiple bounces. In addition, acceleration by $E_\|$ injects electrons \cite{comisso:2019} into the reconnection exhaust at a higher energy and allows the Fermi process to act even more effectively. The detailed partition of energy between the electrons (even if primarily energized by the Fermi process) and ions also depends sensitively on the parallel electric field\cite{haggerty:2015}. 

Over the next decade, a major goal should be understanding how the basic mechanisms of particle energization scale up to large systems, particularly in low-$\beta$ plasmas. Kinetic-scale simulations \cite{shay:2014} and magnetospheric observations \cite{phan:2013,wetherton:2021} found temperature changes $\Delta T$ of each component of the plasma during reconnection scale with the available upstream magnetic energy per particle $\Delta T\propto B^2/2\mu_0n$, suggesting there are large fractional temperature changes in low-$\beta$ plasmas. Similarly, simulation studies find the most significant acceleration of nonthermal particles in low-$\beta$ regimes \cite{li19}. For systems with $\beta_e<\sim0.02$, the effective potential $\phi_\|$  becomes orders of magnitude larger than the thermal energy.  In this case, highly nonlinear density cavities and double layers form \cite{egedal:2015}. This very low-$\beta$ regime is therefore particularly challenging to study because it features superthermal flows, highly anisotropic pressure, and non-adiabatic particle orbits. 

It is not feasible to use brute force kinetic modeling to study very large-scale systems. Rather, reduced models based on physics principles are necessary. Examples may include quasi-neutral kinetic models, hybrid models that combine a kinetic description of some species with a fluid description of others, or extended fluid models \cite{ng:2017}. A promising advance in this vein is a global MHD-scale model that retains the main kinetic acceleration and anisotropy of a population of hot electrons \cite{arnold:2021}. Including the feedback of the pressure anisotropy is important because reconnection heating tends to drive the system towards a state with $\mu_0 (P_\|-P_\perp)\sim B^2$, where pressure anisotropy balances the magnetic tension force that drives the reconnection flows [see Eq.~\eqref{ionMomentum}]\cite{drake06a,yhliu11a,egedal:2013}. Within the next few years, such self-consistent reduced models may be verified against fully kinetic calculations in small systems to determine the range of plasma parameters over which the models are trustworthy and to make predictions for large systems. These models may also move beyond the simple geometries typically considered in fully kinetic simulation. For example, although betatron heating is unimportant in simple current sheet models, it is prevalent in magnetospheric observations of energetic particles at dipolarization fronts \cite{fu:2013,birn:2015}, where plasma drifts back towards Earth's dipole magnetic field. MHD models of contracting magnetic islands in realistic solar flare geometries also show betatron heating can be dominant in compressible systems \cite{borovikov:2017}. Similarly, there is evidence from solar flares that shocks formed as the reconnection outflow jets interact with the background plasma may contribute to substantial particle acceleration \cite{chen:2015}. Betatron and shock heating in the presence of realistic background plasmas and magnetic fields deserve additional theoretical and computational study. 

In conjunction with new computing capabilities (Box 4), improved diagnostics for multiscale experiments and observations (see Box 3) can help solve outstanding problems about reconnection heating and acceleration. Spacecraft missions with more measurement points (satellites) can provide opportunities to understand local energetic particle generation in conjunction with the global reconnection geometry change. Solar observations may provide improved statistics on the energy inventory during flares through either space \cite{mccomas:2019} or ground \cite{chen:2015} missions. And in the laboratory, in addition to in-situ~\cite{fox10} and ex situ~\cite{chien2022} techniques, remote-sensing diagnostics such as fast photodiode detectors~\cite{dubois:2017} show promise for studying energetic particles in large-scale magnetic reconnection experiments. 

In terms of bulk heating, a fundamental unanswered question in solar physics concerns the mechanism by which the solar corona is heated to temperatures hundreds of times greater than the solar photosphere~\cite{klimchuk:2015}. It is becoming accepted that this heating may be from a number of different mechanisms, but the combined heating resulting from numerous small scale reconnection events~\cite{parker72,browning86,hood:2009,priestsyntelis21} is thought to play a crucial role. This is an extremely challenging problem, both for remote sensing observations and for numerical simulations, due to the vast scale separation between the coronal loop sizes and the dissipation scales. With the launch of the Parker Solar probe mission~\cite{velli2020}, a better understanding of the dominant heating mechanisms can conceivably be expected in the near future.

\section*{Outlook}

Many of the central challenges in reconnection physics arise from the large separation between the system scales where magnetic fields become stressed in current sheets and the dissipation scale where these topological constraints are relaxed either through collisional or kinetic mechanisms.   There has been substantial progress in understanding reconnection in fully ionized plasmas where these scales are not too disparate ($\lambda \lesssim 10^2$), including reconnection in the Earth's magnetosphere and small-scale laboratory experiments, both of which are amenable to direct in-situ simulations that resolve kinetic scales.   However, for many solar and astrophysical systems~\cite{ji:2011}, the scale separation is huge ($\lambda \gtrsim 10^8$, $S>10^{15}$), and thus resolving the full range of scales is far beyond foreseeable computing technology.    Additional complications are introduced in partially ionized regimes, which have received far less attention despite of their widespread presence in astrophysics.  Nevertheless, there are promising hypotheses that may explain how fast reconnection works in these large systems, along with exciting opportunities for new progress.  In particular, large-scale current sheets may trigger the MHD plasmoid instability, leading to a hierarchy of flux ropes and new current sheets that extend down to dissipation, often kinetic, scales.  We anticipate that next generation multiscale laboratory experiments, such as FLARE, will critically test this hypothesis in the new regimes of the reconnection phase diagram.  Advanced kinetic simulations will play a key role in interpreting these experiments, including the build-up of free energy and subsequent onset of magnetic reconnection under different driving conditions.  This involves complex coupling between the time evolving current sheet and the larger system, which has widespread applications in nature but too often has been ignored in reconnection studies that start with pre-existing current sheets. As computing advances towards the exascale, we anticipate a great deal of effort on turbulent reconnection in both fluid and kinetic regimes. Can fast reconnection really occur in layers much thicker than kinetic scales, with turbulence controlling the thickness of the diffusion region? This possibility is fundamentally different than regulating the length of the diffusion region by plasmoid formation.  With properly designed numerical and even experimental setups, researchers may begin to distinguish between these two hypotheses in the near future.   

Regardless of which specific scenario is correct, we foresee many exciting research opportunities for understanding heating and nonthermal particle acceleration within these highly dynamic regimes. This includes new approaches for self-consistently including the nonthermal particles within large-scale MHD simulations. Together with novel remote-sensing observations, these approaches may enable progress towards understanding particle acceleration in the solar corona, and perhaps offer constraints on the underlying multiscale physics of magnetic reconnection in large systems. By combining this progress in both in-situ and remote-sensing observations, multiscale simulations and laboratory experiments,  we anticipate that research in the coming years will accelerate towards a more complete understanding of magnetic reconnection in large systems throughout the Universe. 

\section*{Acknowledgements}
HJ, JJA, and JY acknowledge support by the U.S. Department of Energy via Contract No. DE-AC0209CH11466 and WD, AL, and AS acknowledge support by the DOE Frontier Plasma Science Program.  We thank S. Dorfman for providing the top right panel of Fig.2; M. Yamada, A. Bhattacharjee, J. Egedal and F. Guo for their constructive comments on an initial draft of this review. WD acknowledges helpful discussions from the SolFER DRIVE Science Center collaboration.

\section*{Author contributions}
All authors contributed to literature review, drafting text, figure production, discussing, editing, and revising all aspects of this manuscript.

\section*{Competing interests}
The authors declare no competing interests. 

\section*{Peer review information}
Nature Reviews Physics thanks the anonymous reviewers for their contribution to the peer review of this work.

\section*{Publisher’s note}
Springer Nature remains neutral with regard to jurisdictional claims in published maps and institutional affiliations.

\bibliography{refs.bib}
\clearpage


\clearpage
\section*{Box 1: Magnetic reconnection in three dimensions}
\begin{figure*}[t]
\centering{\includegraphics[width=0.5\textwidth]{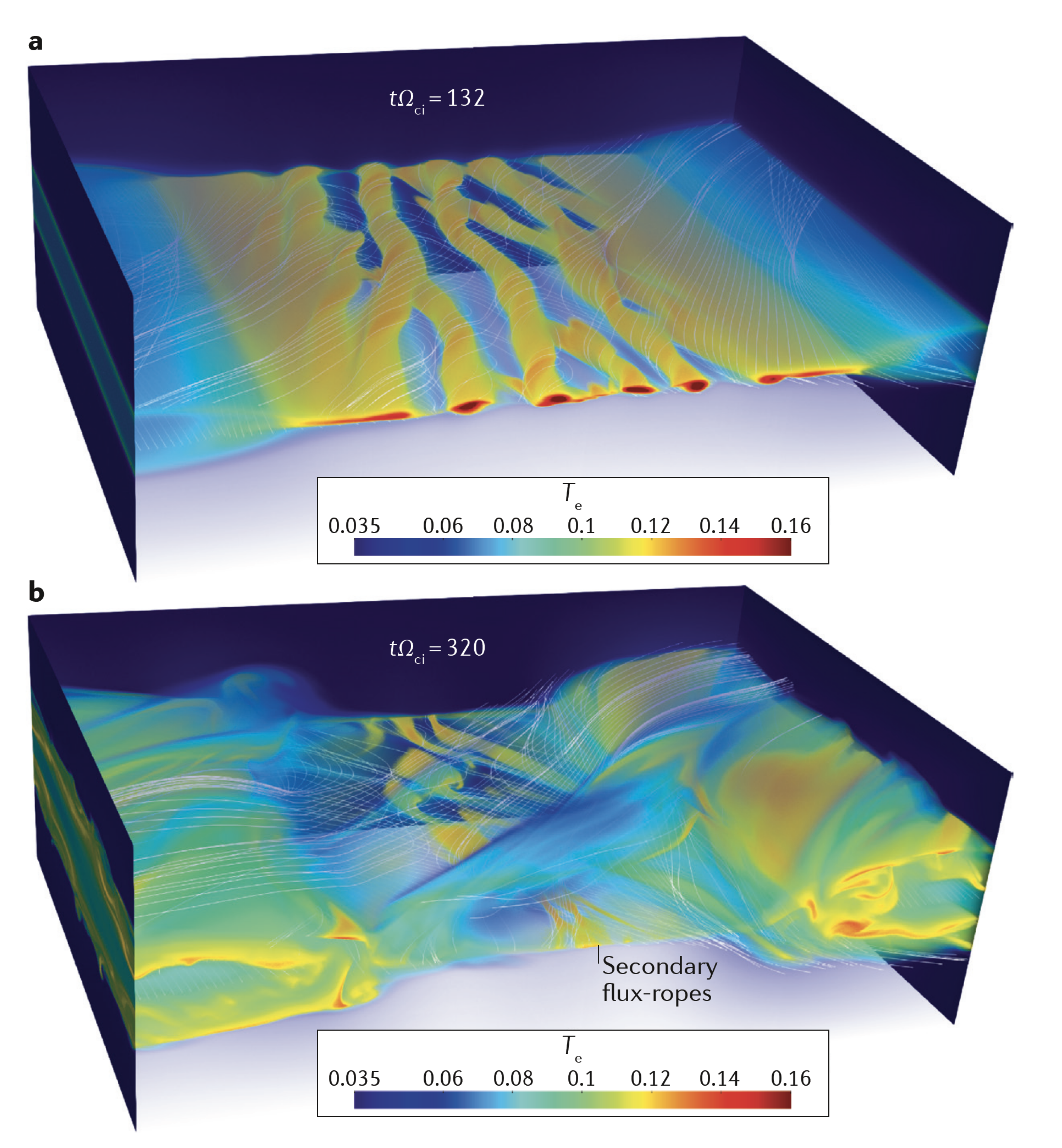}}
\end{figure*}

Understanding of 3D aspects of magnetic reconnection is much less developed~\cite{daughton11a} compared to its 2D counterparts highlighted by the reconnection phase diagram. It is critical to know which features of 2D systems carry over to 3D and which are fundamentally altered. Plasmoids in 2D become flux ropes in 3D which can interact in a complex way by undergoing kink and coalescence instabilities (panel a). Thin kinetic scale current sheets that form between these macro-scale flux ropes grow in length and can be unstable to collisionless tearing modes to form secondary flux-ropes (panel b). Current sheets with a finite guide field have only single resonant surface in 2D, but may have multiples in 3D, which may lead to stochastic field lines through overlapping islands~\cite{rechester78}.  

The required size in the third direction, $L_3$, to study 3D reconnection depends on the subject physics under investigation. For kinetic instabilities, the required minimum $L_3$ is multiples of their wavelengths ranging from Debye scales~\cite{Jara-Almonte2014,fox08}, electron scales (due to beam~\cite{che2011,goldman14} or temperature anisotropy~\cite{kennel66,Yoo18}), lower-hybrid scales~\cite{Carter2001,ji04} (due to cross-field gradient~\cite{krall71} or cross-field drift~\cite{mcbride72}), to ion scales (for drift kink~\cite{daughton99} or kinetic Kelvin-Helmholtz~\cite{nakamura:2017} instabilities), respectively. These kinetic instabilities have attracted interest as potential sources for the enhanced resistivity or viscosity that is often used within fluid descriptions, possibly important in diffusion regions and along the separatrices~\cite{lapenta15} that feature strong spatial gradients and streaming.  

For magnetohydrodynamic (MHD) physics, the required $L_3$ is considerably larger. For example, to permit flux rope kinking~\cite{zhangQile2021}, the minimum $L_3$ is $\pi D (B_{guide}/B_{rec})$ where $D$ is the diameter of the rope for a sufficient twist of field lines.   In collisional MHD regimes, $D \sim L/\sqrt{S_c}$ may be set by the current sheet thickness corresponding to $S_c$. Interaction of kinked and wrapped flux ropes can spontaneously result in complex dynamic structures broadening the current sheet~\cite{huang16} and enhancing dissipation~\cite{wang19}. To avoid recycling information in the periodic systems, the required $L_3 \approx (B_{guide}/B_{rec})L$ is longer if we demand Alfv\'en transit time based on guide field is longer in the third direction than the reconnection time. Fully developed 3D turbulence may become important for reconnection in large systems. In order to avoid recycling information carried by particles, $L_3$ should be longer than their mean free path. This requirement quickly gets stringent especially for nonthermal particles, but is satisfied for a large portion in the operation regime of FLARE to study particle acceleration due to its long circumference.

Lastly, it is still unclear whether an understanding of reconnection physics in periodic systems can be directly applied to natural plasmas, which in general are non-periodic in a 3D geometry, and often line-tied at their ends such as in solar flares. Whether line-tying and driving from the boundaries fundamentally alters reconnection physics has profound importance in connecting laboratory physics, as well as most of numerical research, to astrophysics. A related topic discussed in a series of recent papers, but beyond the coverage of this Roadmap, is about stochastic field lines which can separate rapidly in large 3D turbulent systems~\cite{boozer12a} to lose their connectivity such that reconnection is spontaneously achieved~\cite{eyink11}, but without apparent current sheets.  Magnetic reconnection involves 3D null points is a further advanced subtopic~\cite{priest:2009} whose roles in understanding explosive reconnection is yet to be fully explored~\cite{li21}.

In the figure $T_{\rm{e}}$ is the electron temperature. $t$ is time in the unit of inverse of ion cyclotron angular frequency, $\Omega_{\rm{ci}}$. The figure reproduced with permission from Ref. \cite{stanier:2019}, AIP Publishing.
\section*{Box 2: Open questions}

\vspace{10pt}
Below are some of the major open physics questions regarding magnetic reconnection~\cite{ji2020}:
\begin{itemize}
\item How does reconnection couple global fluid (magnetohydrodynamic) scales to local dissipation (kinetic) scales?
\item How does reconnection take place in three dimensions, in both quasi-2D current sheets (Box 1) and in fully 3D geometries \cite{li21} ?
\item How are particles heated and accelerated?
\item How do boundary conditions affect the reconnection process (Box 1)?
\item How does reconnection start?
\item How does partial ionization affect reconnection?
\item What role does reconnection play in flow-driven systems, which may include self-generated magnetic fields~\cite{hawley92,nilson06}?
\item How does reconnection take place under extreme radiative and relativistic conditions ~\cite{uzdensky11,Zhang:2020}?
\item What role does reconnection play in related processes such as turbulence, shocks and transport (see a recent review~\cite{schekochihin20} and references therein)?
\item How, and under what conditions, is magnetic reconnection a driver or a consequence of explosive phenomena such as Earth's magnetospheric substorms and Coronal Mass Ejections?
\end{itemize}

\section*{Box 3: Multiscale experiments and observations}

Magnetic reconnection in high-temperature, toroidal fusion experiments occurs usually in the multiple X-line collisionless regimes~\cite{ji:2011,tanabe2017} at relatively large Lundquist number $S$ and plasma size $\lambda$, but with limited diagnostics accesses. In contrast, specially implemented diagnostics of many ongoing reconnection experiments in low-temperature linear devices enable detailed studies of 3D electromagnetic fields~\cite{gekelman19,vonStechow16}, quasi-separatrix layers \cite{gekelman16}, electron and ion kinetics~\cite{shi21}, or noninvasive magnetic field measurements~\cite{liu21} at relatively modest $S$ and $\lambda$ values. $S$ or $\lambda$ can be large for the reconnection experiments in high-energy-density (HED) plasmas based on Z pinches~\cite{hare:2017,hare20} or lasers~\cite{nilson06,fiksel14}, but they are driven by converging flows at high $\beta$, in contrast to the typical impulsive reconnection which is driven magnetically at low $\beta$. A unique platform to study particle acceleration by magnetically-driven reconnection at low $\beta$ via ex situ diagnostics is based on capacitor coils powered by lasers~\cite{chien2022}. In addition to FLARE, another upcoming multiscale experiment~\cite{e21} aims to study reconnection in 3D geometries similar to those between shocked solar wind plasma and Earth's magnetosphere.

A common challenge of all these experiments to study multiscale reconnection physics is the lack of a comprehensive set of diagnostics capable of measuring key quantities with sufficient spatial and temporal resolutions, free of severe perturbations to the plasma and restrictive assumptions. The important data include not only multiscale magnetic fields, but also plasma parameters ranging from global images to local kinetic spectra. Progress in understanding multiscale physics critically depends on innovation and efficient implementation of such diagnostics systems in the coming decade.

Challenges and opportunities also exist in both in-situ and remote-sensing multiscale observations~\cite{hesse20}. Building on the current Magnetospheric MultiScale mission~\cite{Burch2016} across election and ion scales, the next-generation multiscale observations require many more satellites to cover larger fluid scales~\cite{kepko18,klein21}. However, specific strategies to efficiently capture multiscale reconnection physics, such as usage of data science techniques~\cite{sitnov21}, need to be developed and tested based on numerical simulations as well as observational and laboratory data. Alternative approaches, such as imaging at ultraviolet and soft X-ray wavelengths of Earth's magnetosphere~\cite{SMILE21}, may provide much needed information on the global consequences of magnetic reconnection.

The Solar Parker Probe (PSP), Solar Orbiter, and Daniel K. Inouye Solar Telescope (DKIST) in addition to the existing Solar Dynamics Observatory (SDO) and Interface Region Imaging Spectrograph (IRIS) have opened new ways of solar reconnection research such as mapping field line connectivity from in-situ measurements at close distances from the Sun to high-resolution images of solar surface and corona. Rapid imaging at microwave and radio frequencies from the Expanded Owens Valley Solar Array (EOVSA), and its next generations, has enabled measurements of spatially resolved magnetic field and relativistic electrons~\cite{chen:2020} as a direct product of multiscale reconnection during flares.

\section*{Box 4: Multiscale simulations and exascale computing\label{s:algorithms}}

Despite considerable progress, simulations remain challenging due to the extreme disparity between the global and kinetic scales. In many problems, magnetohydrodynamics (MHD) remains the only feasible option whereas kinetic simulations are used for detailed studies of local regions. Progress towards bridging these scales will come from the following advances in simulation.
\begin{itemize}
\item{Exascale computing.} At present, MHD simulations with $S \sim 10^6$ and $\sim10^9$ cells~\cite{dong:2018, yang:2020} and kinetic particle-in-cell (PIC) simulations with $L/d_i \lesssim 10^2$ corresponding to $10^9$ cells and $10^{12}$ particles~\cite{daughton11a,guo:2020,stanier:2019} are currently possible.  Exascale computing is expected to expand these capabilities, but will also involve a range of new challenges (optimizing for new architectures, storage, visualization and analysis of petabye data sets). These challenges are cross-disciplinary, and can be addressed with high-performance libraries. For example, the PIC code VPIC has been integrated with the Kokkos performance portability library~\cite{bird:2021} achieving a performance of $2\times 10^{9}$ particle pushes/second/GPU on V100 GPUs. For computers approaching the exascale (Summit, 200 petaflops - 27,648 V100 GPUs) this will permit $55\times10^{12}$ particle pushes/sec.   Thus, we anticipate that kinetic simulations with $\sim 5\times 10^{10}$ cells, and $\sim 10^{13}$ particles will soon be possible.  Simulations that were heroic at the petascale, will become routine at the exascale, allowing researchers to perform scaling studies to understand key dependencies.   

\item{Model reduction.} Even at the exascale, global fully kinetic simulations will remain infeasible for most problems. Reduced models are needed of higher-fidelity than MHD, but with the stiffest kinetic scales removed. One possibility is extended MHD approaches using a truncated hierarchy of moment equations, with approximate closures that may be sufficiently accurate to model reconnection~\cite{le:2009,ng:2017,allmann:2018}. Exciting future directions include machine learning~\cite{maulik:2020}, and sub-grid models for reconnection turbulence~\cite{widmer:2016}. Hybrid models combine the fluid approach with some plasma components that are treated kinetically, such as the kinetic-ion fluid-electron~\cite{winske:2003,karimabadi:2004} or fast-particle-kinetic bulk-MHD models~\cite{arnold:2021}. Finally, the gyro-kinetic/fluid models have shown promise for modeling reconnection in strong guide field regimes~\cite{loureiro:2013, tenbarge:2014}.  To understand the strengths and limitations of these various approaches, cross comparisons have proven invaluable ~\cite{birn01a,karimabadi:2004,tenbarge:2014,stanier:2015,ng:2015,ng:2017}.
 
\item{Advanced algorithms.} There are a variety algorithmic advances that show promise for reconnection physics. Advanced spatial discretizations and mesh refinement have shown excellent results for resolving thin reconnection layers in MHD simulation~\cite{karpen:2012,delsarto:2017}, and their use is becoming more widespread in kinetic and reduced models. Implicit methods can step over stiff timescales in a stable manner, sometimes with superior numerical conservation properties~\cite{markidis2011,chen2011,stanier2019hybrid}. Finally, there have been exciting advances in coupling different fidelity models using domain decomposition. The embedded-PIC method~\cite{makwana:2018} uses an MHD or extended-MHD fluid model on large scales, and couples it to localized PIC boxes at the diffusion regions.
\end{itemize}

\clearpage

\section*{Glossary terms}

\textbf{collisional plasma}:  (or collisionles) plasmas in which Coulomb collisions are important(unimportant) for the subject of interest, which in this Roadmap is magnetic reconnection.\\ 

\textbf{magneto-rotational instability}: a plasma instability believed to generate turbulence to explain the observed fast accretion in magnetized astrophysical disks where angular speed increases while specific angular momentum decreases radially.\\

\textbf{Kelvin-Helmholtz instability}: a linear instability driven by velocity shear in a fluid or plasma.\\

\textbf{magnetic Prandtl number}: a dimensionless parameter in magnetic hydrodynamic fluids or plasmas to quantify the momentum diffusion relative to magnetic diffusion.\\

\textbf{kink instability}: a plasma instability which produces helical kinking of a current channel and is driven by excessively large electric currents for a given magnetic flux in the same direction.\\

\textbf{ballooning instability}: a plasma instability which causes the magnetic field to balloon outwards towards the weak field direction due to excessively large plasma pressure gradient.\\

\textbf{torus instability}: An expansion magnetic hydrodynamic instability of current carrying torus in solar and laboratory plasmas due to rapid decrease of the required equilibrium transverse magnetic field in the expansion direction.\\

\textbf{sawtooth oscillation}: A quasi-periodic, sawtooth-like oscillation in soft X-ray measurements of the core tokamak plasmas from rapid loss and gradual recovery of hot electron temperature due to an internal magnetic hydrodynamic instability.\\

\end{document}